\documentclass[useAMS,usenatbib]{mnras}
\usepackage{amsmath}
\usepackage{graphicx,txfonts,fleqn,enumerate,color,bm}
\usepackage{color}
\usepackage{ulem}

\renewcommand*{\v}[1]  {\boldsymbol{#1}}

\newcommand*  {\twovector}[2] {{\begin{pmatrix} $1 \\ $2 \end{pmatrix}}}

\renewcommand {\emph}[1]  {\textit{#1}}

\title[Time{-}symmetric integration in astrophysics]
{Time-symmetric integration in astrophysics}

\author[David M. Hernandez and Edmund Bertschinger]
	{David M. Hernandez \thanks{Email: dmhernan@mit.edu (DMH); edbert@mit.edu (EB)} and Edmund Bertschinger \footnotemark[1]  \\
	Department of Physics and Kavli Institute for Astrophysics and Space Research,
	Massachusetts Institute of Technology, 77 Massachusetts Ave., \\ Cambridge, Massachusetts 02139, USA\\
	}
	
\begin{document}

\maketitle

\label{first page}
\begin{abstract}
Calculating the long term solution of ordinary differential equations, such as those of the $N$-body problem, is central to understanding a wide range of dynamics in astrophysics, from galaxy formation to planetary chaos.  Because generally no analytic solution exists to these equations, researchers rely on numerical methods which are prone to various errors.  In an effort to mitigate these errors, powerful symplectic integrators have been employed.  But symplectic integrators can be severely limited because they are not compatible with adaptive stepping and thus they have difficulty accommodating changing time and length scales.  A promising alternative is time-reversible integration, which can handle adaptive time stepping, but the errors due to time-reversible integration in astrophysics are less understood.  The goal of this work is {to} study analytically and numerically the errors caused by time-reversible integration, with and without adaptive stepping.  We derive the modified differential equations of these integrators to perform the error analysis.  As an example, we consider the trapezoidal rule, a reversible non-symplectic integrator, and show it gives secular energy error increase for a pendulum problem and for a H\'{e}non{--}Heiles orbit.  We conclude that using reversible integration does not guarantee good energy conservation and that, when possible, use of symplectic integrators is favored.  We also {show} that time-symmetry and time-reversibility are properties that are distinct for an integrator.
\end{abstract}
\begin{keywords}
methods: numerical{---}celestial mechanics{---}globular clusters: general{---} planets and satellites: dynamical evolution and stability{---} galaxies
\end{keywords}
\section{Introduction} 
Obtaining solutions to initial value problems of ordinary differential equations (ODEs) over long time periods is central to dynamical calculations in astrophysics.  These ODEs might represent problems such as the $N$-body problem, $N$ point particles interacting through pairwise forces, or the problem of particle orbits in a time-independent galactic potential.  The ODEs are frequently described by a time-dependent or time-independent Hamiltonian.  

Obtaining a solution to the $N$-body problem is essential for many purposes, from calculating the evolution of dark matter in the Universe to understanding stability and chaos of orbits in planetary systems.  Different techniques, relying on different assumptions, have been developed to obtain approximate $N$-body solutions.  The $N$-body problem is generally chaotic and non-integrable, so we rely on these approximations to obtain its solutions.  But, in general $N$-body cases, it is unknown how reliable such approximations are.  In fact, the approximations themselves give rise to chaos, separate from the physical chaos of the problem itself.  If the numerical method itself can be responsible for chaos, then the error from the original trajectory can grow exponentially, leading to call into question the validity of the calculated solution.

Galactic potentials usually have only a few degrees of freedom, but can still be chaotic and non-integrable, and suffer from the same problems described above.  In fact, much of the study of chaos began with the study of the H\'{e}non{--}Heiles problem, which was motivated by the study of galactic potentials.  

It would appear numerical approximation to chaotic ODEs should be suspect, but fortunately, geometric numerical integration, integration aimed at respecting the geometry of the underlying ODEs, has been developed and helps restore confidence in these numerical solutions \citep{chan90}.  Depending on the equations, geometric properties include the Hamiltonian flow, time-reversibility, and quadratic and linear invariants in the phase space.  In the last 30 years, astrophysics researchers have made geometric integration a standard in various fields of dynamics, including planets \citep{WH91,C99,DLL98,H16}, stellar clusters \citep{K98,HMM95,HB15,DH17}, or galaxy formation \citep{S05}.  

Geometric integrators that respect Hamiltonian flow are also called symplectic integrators, and they conserve generalizations of volumes in phase space, also known as Poincar\'{e} invariants.  The theory of symplectic integration is well developed \citep{hair06}. The citations above \citep{WH91,C99,DLL98,H16,K98,HMM95,HB15,DH17,S05}, are all concerned with time-independent Hamiltonian problems, so a symplectic integrator is ideal.  However, symplectic integrators applied to the above problems have a limitation; if the step sizes are chosen as a function of the phase space, the evolution of the trajectory is no longer Hamiltonian.  This limitation is severe for the $N$-body problem because gravity has no length scale: two-body relaxation is affected by close and far encounters.  Thus, the range of time and length scales is large, posing a severe challenge for fixed time step integration.

Thus, some researchers \citep{pelu12,HMM95,K98} have abandoned the requirement of a symplectic integrator and instead focused on integrators that preserve time-reversibility, if the underlying equations have this symmetry.  It is important to note there exist irreversible conservative differential equations{---} see Section \ref{sec:mde}.  However, many important problems such as the $N$-body problem are conservative and reversible.  Time-reversible integration appears to be less studied than symplectic integration, but it has been observed that time-reversible integrators generally can reduce errors for integrations in astrophysics.  An explanation for such behavior is sometimes not provided.  It is possible to adapt time steps while still preserving time-reversibility \citep{Makinoetal2006,Funatoetal96}, so clearly we would like to abandon symplecticity if possible and if time-reversible integration is good enough.  But a clear error analysis is needed with these integrators in order for one to have confidence in their use.

The goal of this paper is to provide that error analysis, and to use it to show that the behavior of a time-reversible integrator in astrophysics can be worse than a symplectic integrator.  This suggests that researchers in astrophysics should use symplectic integrators when possible.  To perform the error analysis, we derive the modified differential equations (MDEs) obeyed by these methods using adaptive time steps, and we use these equations to calculate how well the methods conserve energy.  We study a simple pendulum problem and H\'{e}non{--}Heiles orbits.  We show how various reversible integrators do not conserve energy to all orders.  It was already noted by \cite{FHP04} that some fixed-step Runge{--}Kutta reversible methods do not conserve an energy.

Section \ref{sec:rev} shows the tools necessary for deriving the MDE.  Section \ref{sec:trap} derives the MDE for the trapezoidal rule, a non-symplectic but symmetric second order Runge{--}Kutta method.  We also derive the MDE for the trapezoidal rule with adaptive steps.  We derive various properties of the trapezoidal rule.  In Section \ref{sec:numdem}, we apply our numerical analysis to understand the error in energy of the modified pendulum problem and H\'{e}non{--}Heiles orbits and to find that time-symmetric integration can give energy drift.  While the analysis in this section is limited to the trapezoidal rule, there is no reason to believe other time symmetric integrators would not suffer from energy drift.  In the Appendices, we show that for Runge{--}Kutta methods, time-symmetry, reversibility, and symplecticity are independent concepts.  We conclude in Section \ref{sec:conc}.

\section{The modified differential equation} 

\subsection{Time-symmetric integration} \label{sec:rev}
A system of autonomous ordinary differential equations can be written
\begin{equation}\label{sysode}
  \dot{\v y}={\v f}({\v y})
\end{equation}
where ${\v y}$ and ${\v f}$ are both vectors of length $n$. We are concerned with the case where $\v{y}$ is a vector of positions and velocities or the phase space defined by canonical coordinates and momenta: in equations, $\v{y} = (\v{q},\v{v})$ or $\v{y} = (\v{q},\v{p})$.  The problem is to find ${\v y}(t)$ given ${\v y}(0)\equiv{\v y}_0$. We assume that the system is autonomous, so that ${\v f}$ depends only on ${\v y}$ and not on $t$.

A numerical one-step method estimates the solution at $t=h$, ${\v y}(h)\approx{\v y}_1$, where
\begin{equation}\label{algor}
  {\v y}_1={\v y}_0+h{\v G}({\v y_0},h)
\end{equation}
for some ${\v G}$ that is related to ${\v f}$. The method may be iterated to estimate the solution at $t=2h,\,3h,$ etc. For now, we assume that $h$ is a constant independent of ${\v y}_0$ and $t$.

A goal of numerical analysis is to find a ${\v G}$ that is inexpensive to evaluate so that $|{\v y}_1-{\v y}(h)|$ is smaller than a specified tolerance. {The study of the errors $|{\v y}_1-{\v y}(h)|$ is known as forward error analysis.}  One can turn around the problem. Given ${\v G}$, find a modified differential equation whose exact solution is ${\v y}(h)={\v y}_1$. That modified differential equation is written
\begin{equation}\label{MDE}
  \dot{\v y}={\v F}({\v y},h)\ .
\end{equation}
The goal then becomes to minimize {$|{\v F}({\v y},h)-{\v f}({\v y})|$}. This is done by determining ${\v F}({\v y},h)$ from ${\v G}({\v y},h)$.  {Studying the errors $|{\v F}({\v y},h)-{\v f}({\v y})|$ is known as backward error analysis.}

A symmetric one-step integrator is one for which a forward step $h$ followed by a backward step $-h$ restores the initial conditions. The requirement is
\begin{equation}\label{reverse}
  {\v G}({\v y}_1,-h)={\v G}({\v y}_0,h)=\frac{1}{h}
    ({\v y}_1-{\v y}_0)\ .
\end{equation}
The associated modified differential equation is even: ${\v F}({\v y},-h)={\v F}({\v y},h)$

$\rho${-}reversibility \citep[Section V.1]{hair06} means that if we change the sign of velocities, while keeping the position coordinates constant, the solution trajectory must stay the same{---} only the direction of motion is inverted.  Let $\bm{\rho}$ be an invertible linear transformation that changes the signs of velocities: $\bm{\rho} \v{y} = \bm{\rho} (\v{q},\v{v}) = (\v{q},-\v{v})$.  All autonomous Newtonian physics problems are described by positions and velocities and can be written as a system of first order ODE's: $\dot{\v{q}} = \v{f}(\v{q},\v{v})$ and $\dot{\v{v}} = \v{g}(\v{q},\v{v})$.  They are not all reversible; if they are, then,
\begin{equation}
\begin{aligned}
\v{f}(\v{q},-\v{v}) &= -\v{f}(\v{q},\v{v}), \qquad \text{and} \\
\v{g}(\v{q},-\v{v}) &= \v{g}(\v{q},\v{v}).
\label{eq:rev}
\end{aligned}
\end{equation}
While many problems satisfy this requirement, not all do.  For example, the system of differential equations for a charged particle moving in a magnetic field are
\begin{equation}
\begin{aligned}
\dot{\v{v}} &= \frac{e}{m} \left(\v{v} \times \v{B}(\v{q},t)\right), \qquad \mathrm{and} \\
\dot{\v{q}} &= \v{v},
\end{aligned}
\end{equation}
where $e$ is the charge of the particle, $m$ is the mass of the particle, and $\v{B}$ is the external magnetic field.
These equations do not satisfy \eqref{eq:rev}; the solution trajectory is different when we switch the sign of the velocities {(the resolution of this apparent irreversibility is that the sign of $\v{B}$ changes if we reverse the currents causing it)}.  

If we use a one-step method to solve a $\rho$-reversible set of differential equations, the symmetric integrator is $\rho$-reversible.  The $\rho$-reversibility condition for an integrator is connected to \eqref{eq:rev}:
\begin{equation}
\label{eq:revsymm}
\bm{\rho} \phi_h \v{y} = \phi_{-h} \bm{\rho} \v{y},
\end{equation}
which implies $\phi_h \bm{\rho} \phi_h = \phi_h \phi_{-h} \bm{\rho}$.  This only holds if the integrator is time-symmetric, or $\phi_h \phi_{-h} = I$.  Thus, in what follows, until Section \ref{sec:modad}, `symmetric' one-step methods will be equivalent to`time-reversible' one-step methods because we are only concerned with $\rho$-reversible differential equations.  However, it is important to bear in mind that a symmetric method is not necessarily the same as a time-reversible method; one way to break the equivalency is by letting the step $h$ vary as a function of phase space.

\subsection{Derivation of modified differential equation} \label{sec:mde}
Our goal is to understand time-symmetric integrators.  Unlike symplectic methods, symmetric integrators {generally}  have no surrogate Hamiltonian \citep[Section IX.8]{hair06} which informs us of the dynamics, so we instead derive the differential equations the integrator obeys.  We call this the modified differential equations (MDEs), and its study has been referred to as backward error analysis \citep[Chapter IX]{hair06}.  

Proceed as follows: first write the formally exact solution of equation (\ref{MDE}) with initial condition ${\v y}={\v y}_0$,
\begin{equation}\label{MDEsol}
  {\v y}_1=\exp(h\tilde D){\v y}_0={\v y}_0+\sum_{n=1}^\infty
    \frac{h^n}{n!}\tilde D^{n-1}{\v F}({\v y}_0,h)\ ,\ \ 
 \tilde D\equiv{\v F}({\v y}_0,h)\cdot\frac{\partial}{\partial{\v y}_0}\ .
\end{equation}
{This is just the usual Taylor expansion solution.}  Next expand ${\v F}({\v y}_0,h)$ and ${\v G}({\v y}_0,h)$ in power series in $h$:
\begin{equation}\label{TaylorFG}
  {\v F}({\v y}_0,h)=\sum_{n=0}^\infty h^n{\v f}_n({\v y}_0)\ ,\ \ 
   {\v G}({\v y}_0,h)=\sum_{n=0}^\infty h^n{\v g}_n({\v y}_0)\ .
\end{equation}
{$\v{f}_0$ is the $\v{f}$ from \eqref{sysode}.}  Use (\ref{TaylorFG}) to expand the derivative operator
\begin{equation}\label{Dexp}
  \tilde D=\sum_{n=0}^\infty h^nD_n\ ,\ \ 
  D_n\equiv{\v f}_n({\v y}_0)\cdot\frac{\partial}{\partial{\v y}_0}\ .
\end{equation}
Combining equations (\ref{algor}) and (\ref{MDEsol})--(\ref{Dexp}) gives, for $n\le4$,
\begin{eqnarray}\label{gf}
  {\v g}_0&=&{\v f}_0\nonumber\\
  {\v g}_1&=&{\v f}_1+\frac{1}{2}D_0{\v f}_0\nonumber\\
  {\v g}_2&=&{\v f}_2+\frac{1}{2}(D_0{\v f}_1+D_1{\v f}_0)
    +\frac{1}{6}D_0^2{\v f}_0\nonumber\\
  {\v g}_3&=&{\v f}_3+\frac{1}{2}(D_0{\v f}_2+D_1{\v f}_1
    +D_2{\v f}_0)+\frac{1}{6}(D_0^2{\v f}_1+D_0D_1{\v f}_0
    +D_1D_0{\v f}_0)\nonumber \\
    &&\qquad\ + \frac{1}{24}D_0^3{\v f}_0\nonumber\\
  {\v g}_4&=&{\v f}_4+\frac{1}{2}(D_0{\v f}_3+D_1{\v f}_2
    +D_2{\v f}_1+D_3{\v f}_0)\nonumber\\
  &&\quad\ +\,\frac{1}{6}(D_0^2{\v f}_2+D_0D_2{\v f}_0
    +D_2D_0{\v f}_0+D_0D_1{\v f}_1+D_1D_0{\v f}_1+D_1^2
    {\v f}_0)\nonumber\\
  &&\quad\ +\,\frac{1}{24}(D_0^3{\v f}_1+D_0^2D_1{\v f}_0
    +D_0D_1D_0{\v f}_0+D_1D_0^2{\v f}_0)\nonumber \\
    &&\quad + \frac{1}{120}
    D_0^4{\v f}_0\ .
\end{eqnarray}

Our goal is to obtain ${\v F}$ from ${\v G}$. One way is to solve equations (\ref{gf}) recursively, starting with ${\v f}_0={\v g}_0$ substituting into ${\v f}_1={\v g}_1-\frac{1}{2}D_0f_0$, and so on. This is useful for determining ${\v f}_n$ for small $n$.

As an example, consider the explicit Euler method
\begin{equation}\label{EEuler}
  {\v y}_1={\v y}_0+h{\v f}({\v y}_0)\ ,
\end{equation}
for which ${\v g}_0={\v f}$, ${\v g}_1={\v g}_2={\v g}_3=0$. This method is first order because
${\v F}({\v y},h)={\v f}({\v y})-\frac{1}{2}hD_0{\v f}({\v y})+O(h^2)$. For a{n} $n$th order method,
\begin{equation}\label{nthorder}
  {\v g}_k=\frac{1}{(k+1)!}D_0^k{\v f}\ ,\ \ 0\le k\le n-1\ .
\end{equation}

Recursive solution is impractical to extend to high order.   An alternative approach (which may also be difficult, but is conceptually appealing) is to sum the series for ${\v F}$ in equation (\ref{MDEsol}) by defining the differential operator
\begin{equation}\label{Gder}
  \tilde G({\v y}_0,h)\equiv{\v G}({\v y}_0,h)\cdot\frac{\partial}
    {\partial{\v y}_0}.
\end{equation}
Then
\begin{equation}\label{Fsol}
  h\tilde D=\ln(1+h\tilde G)=h\tilde G-\frac{1}{2}(h\tilde G)^2+\frac{1}{3}
    (h\tilde G)^3-\frac{1}{4}(h\tilde G)^4+\cdots\ .
\end{equation}
The logarithm of an operator is defined by its series expansion. Applying the operators to ${\v y}_0$ gives ${\v F}({\v y}_0,h)=\tilde D
{\v y}_0$ and $\tilde G{\v y}_0={\v G}({\v y}_0,h)$ so that
\begin{equation}\label{Fser}
\begin{aligned}
  {\v F}&={\v G}-\frac{1}{2}h\tilde G{\v G}+\frac{1}{3}h^2
    \tilde G^2{\v G}-\frac{1}{4}h^3\tilde G^3{\v G}+\frac{1}{5}
    h^4\tilde G^4{\v G}-\cdots\ \\
    &= (h \tilde{G} )^{-1} \ln (1 + h \tilde{G} ) \v{G}.
    \end{aligned}
\end{equation}
{The relation $\tilde{G}^{-1} \tilde{G} = 1$ defines $\tilde{G}^{-1}$.}

\section{A study of the trapezoidal rule: a time-symmetric but non-symplectic integrator}
\label{sec:trap}

\subsection{Relating the trapezoidal and midpoint rule}
\label{sec:rel}
 We introduce {several} one-step integrators.  Let $\phi_h^T, \phi_h^M, \phi_h^E$, and $\phi_h^I$ indicate the trapezoidal, midpoint, explicit Euler, and implicit Euler one-step integration methods, respectively.  The midpoint rule is symplectic, while the trapezoidal rule is not, but they have a close connection.  The two integrators are defined by 
\begin{equation}
\phi_h^T \v{y}_0 = \v{y}_1 = \v{y}_0  + \frac{h}{2} \left[ \v{f}(\v{y}_0) + \v{f}(\v{y}_1)\right], 
\label{eq:trap}
\end{equation}
and
{
\begin{equation}
\phi_h^M \v{y}_0 = \v{y}_1 = \v{y}_0  + h \v{f}\left(\frac{\v{y}_0 +\v{y}_1}{2}\right).
\label{eq:mid}
\end{equation}
}
\noindent The explicit and implicit Euler methods are first-order, not time-symmetric, and non-symplectic.  They are
\begin{equation}
\begin{aligned}
\phi_h^E \v{y}_0 &= \v{y}_1 = \v{y}_0  + h \v{f}(\v{y}_0) , \qquad \mathrm{and}\\
\phi_h^I \v{y}_0 &= \v{y}_1 = \v{y}_0  + h \v{f}(\v{y}_1).\\
\end{aligned}
\end{equation}

We see that
\begin{equation}
\begin{aligned}
\phi_h^T = \phi_{h/2}^I \phi_{h/2}^E, \qquad \mathrm{and} \qquad \phi_h^M = \phi_{h/2}^E \phi_{h/2}^I,
\end{aligned}
\end{equation}
so that $\phi_h^T = (\phi_{h/2}^E)^{-1} \phi_h^M \phi_{h/2}^E$.  Thus, the trapezoidal and midpoint rules are said to be {\it conjugate} \citep[Section VI.8]{hair06} to each other.  To get a trapezoidal orbit, we need only apply a correction at the beginning and ending of a midpoint rule integration.  This means the trapezoidal rule solution should have similar error properties to a symplectic method like the midpoint rule; we will show this more carefully in Section \ref{sec:consq}.  
\subsection{Runge{--}Kutta methods}
\label{sec:RK}

The numerical algorithm (\ref{algor}) is a mapping of the vector space $\{{\bm y}\}$ onto itself. A broad class of integrators, that encompasses various common algorithms including the ones of Section \ref{sec:rel}, defines the mapping ${\bm y}_0\to{\bm y}_1={\bm y}_0+h{\bm G}({\bm y}_0,h)$ using only ${\bm f}({\bm y}_0)$ and derivative operators that are scalars under coordinate transformations of ${\bm y}$.  They are called Runge{--}Kutta (RK) methods.  An RK method of $s$ stages is defined by constants $a_{i j}$ and  $b_i$ for $1 \le i,j \le s$ {when there is no explicit time-dependence in the governing ODE's}:{
\begin{equation}
\label{eq:RK}
\begin{aligned}
\v{y}_1 &= \v{y}_0 + h \sum_{i = 1}^s b_i \v{k}_i, \qquad \mathrm{and}\\
\v{k}_i &= \v{f}(\v{y}_0 + h \sum_{j = 1} ^s a_{i j} \v{k}_j),
\end{aligned}
\end{equation} }
\noindent which is explicit, and thus less computationally expensive, if and only if $a_{i j} = 0$ for $j \ge i$ (a strictly triangular matrix).  For this method, ${\bm G}({\bm y},h)$ depends on ${\bm f}$ and on differential operators like $D_0\equiv{\bm f}({\bm y})\cdot(\partial/\partial{\bm y})$ that are scalars under general coordinate transformations ${\bm y}\to{\bm y}'$.  The popular leapfrog method is not a{n} RK method{---}it is known as a partitioned Runge{--}Kutta method{---} because it uses a different rule for updating the positions and momenta.  {In the partitioned RK case,} the differential operators defining the ${\bm g}_k$ are no longer covariant under general linear transformations of the full space.  

The methods \eqref{eq:trap} and \eqref{eq:mid} are RK methods.  We can check that for the former, $s=2$, $b_1 = b_2 = 1/2$ , $a_{21}=a_{22} = 1/2$, and $a_{11}=a_{12} = 0$.  For the latter, $s=1$, $a_{11} = 1/2$, and $b_1 = 1/2$.  Both are implicit and thus will need to be solved through iteration, whether fixed-point or {Newton-Rhapson} \citep{press02}.

It is easy to see both the implicit midpoint {(as opposed to explicit midpoint, a different RK method)} and trapezoidal rule are time-symmetric (and reversible, cf. Section \ref{sec:rev} ) {\bf if} used with fixed time step. Take a step forwards from $\v{y}_0$ to obtain $\v{y}_1$ and a step backwards to obtain $\v{y}^\prime$.  The rules require $\v{y}^\prime = \v{y}_0$.  

Next, we investigate whether the methods are symplectic.  It has been shown \citep[Section VI.4]{hair06}, that if and only if a{n} RK method conserves quadratic invariants in the phase space variables of the underlying differential equations, it is symplectic.  The reason is related to the fact that {the functions of $\v{y}$, $\bm{S J S}^\dagger$, defined by equations \eqref{SympJacob}, which must be invariant for symplecticity to hold, are first integrals} of the variational equations.  One such typical quadratic invariant in some problems is the angular momentum.  Any quadratic invariant can be written $Q(\v{y}) = \v{y}^{\dagger} \bm{C} \v{y}$, with $\bm{C}$ a symmetric matrix.  Write the implicit midpoint rule as
\begin{equation}
\v{y}_1 - \v{y}_0 = h \v{f}\left(\frac{\v{y}_1 + \v{y}_0}{2} \right).
\end{equation}
Multiply from the left by $\left(\v{y}_1 + \v{y}_0\right)^\dagger \bm{C}$.  {The left hand side gives
\begin{equation}
\v{y}_1^\dag \bm{C} \v{y}_1 - \v{y}_1^\dag \bm{C} \v{y}_0 + \v{y}_0^\dag \bm{C} \v{y}_1 - \v{y}_0^\dag \bm{C} \v{y}_0 = \v{y}_1^\dagger \bm{C} \v{y}_1 - \v{y}_0^\dagger \bm{C} \v{y}_0,
\end{equation}
which follows from the fact that the transpose of a scalar is the scalar.  The right hand is zero because $\dot{Q}(({\v{y}_1 + \v{y}_0})/{2}) = 0$.  Thus, }we are left with $\v{y}_1^\dagger \bm{C} \v{y}_1 - \v{y}_0^\dagger \bm{C} \v{y}_0 = 0$,
which means {the implicit midpoint rule} conserves quadratic invariants and is thus symplectic.  Any numerical experiment with a symplectic integrator that is a{n} RK method will show conservation of all quadratic invariants; an example is the angular momentum, for differential equations that have this symmetry, such as the Kepler problem.  We show a more direct proof of the symplecticity of the midpoint rule in Appendix \ref{sec:symp}.

Write the trapezoidal rule as
\begin{equation}
\label{eq:trapprime}
\v{y}_1 - \v{y}_0 = \frac{h}{2} \left[\v{f} (\v{y}_0) + \v{f}(\v{y}_1) \right].
\end{equation}
If we multiply on the left by $(\v{y}_1 + \v{y}_0)^\dagger \bm{C}$, we find that $\v{y}_1^\dagger \bm{C} \v{y}_1 - \v{y}_0^\dagger \bm{C} \v{y}_0 \ne 0$ and is generally not conserved, meaning quadratic invariants are not conserved, and the trapezoidal rule is not symplectic.  Numerical experiments indeed show the trapezoidal rule does not conserve quadratic invariants such as the angular momentum.

We will largely  focus on the trapezoidal rule for the remainder of the paper, because we are interested in a time-symmetric, but non-symplectic integrator, and this method is one of the simplest examples of this.  Some researchers have used leapfrog, which is symplectic {when using fixed time step}, with reversible steps.  Once the steps are adapted, however, the symplectic property is lost, so there is no advantage from this standpoint to use leapfrog.  On the other hand, even when used with adaptive steps, leapfrog conserves angular momentum exactly, while the trapezoidal rule does not.  However, the tests we consider in what follows have no angular momentum invariant.  Also, the trapezoidal rule has a related invariant for every quadratic, in the phase space, invariant in the underlying equations; see Section \ref{sec:consq}.  Both leapfrog and trapezoidal rule conserve linear invariants, such as the total linear momentum, exactly (all RK methods do).  {A disadvantage of the trapezoidal rule is that it requires solving implicit equations, unlike leapfrog.  But any time-symmetric Runge--Kutta method is implicit.  We focus our efforts on Runge--Kutta methods because their properties have been well established, and they treat all phase space components with the same functional update rule, which will simplify our analysis of their symplecticity and energy conservation properties in Appendices \ref{sec2} and \ref{sec3}.  An advantage of the trapezoidal rule, as shown in Section \ref{sec:rel}, is its connection to a symplectic method.}

\subsection{A conserved quantity for the trapezoidal rule}
\label{sec:consq}
Consider the broad class of separable Hamiltonians, 
{
\begin{equation} \label{eq:broad}
H_0 = \sum_{i=1}^n \frac{p_i^2}{2 m_i} + U(\v{q}),
\label{eq:h0}
\end{equation}
}
\noindent {where $2n$ is the phase space dimension,} and define 
\begin{equation}
U_i \equiv \frac{\partial U}{\partial q_i}.
\end{equation}
{Additional derivatives of $U$ with respect to the $q_i$ are denoted by more $U$ subscript indices.  }{In Section \ref{sec:trap}, l}et $\v{y}^{\mathrm{T}} = (\v{q}^{\mathrm{T}},\v{p}^{\mathrm{T}})$ and $\v{y}^{\mathrm{M}} = (\v{q}^{\mathrm{M}},\v{p}^{\mathrm{M}})$ refer to $\v{y}_1$ from the trapezoidal and implicit midpoint rule, respectively.   {Let other functions be functions of $\v{y}_0$.  As an example, $q_i^T = q_i + h p_i - h^2/2 U_i + \mathcal O(h^3)$

}

In Appendix \ref{sec:MDEMT}, we derive the modified differential equations for the trapezoidal and implicit midpoint rule, and the Hamiltonian for the implicit midpoint rule.  Using the results from {Appendix} \ref{sec:MDEMT}, we find
{
\begin{equation}
\begin{aligned}
q_i^\mathrm{T} &=  q_i^\mathrm{M} + \mathcal O(h^4), \\
p_i^\mathrm{T} &= p_i^\mathrm{M} - \frac{h^3}{8} \sum_{j,k = 1}^n \frac{p_j p_k}{m_j m_k} U_{i j k} + \mathcal O(h^4), \\
\dot{q}_i^\mathrm{T} &=  \frac{\partial \tilde{H} }{\partial {p_i}}  + \mathcal O(h^4), \\
\dot{p}_i^\mathrm{T} &= - \frac{\partial \tilde{H}}{\partial {q_i}}  - \frac{h^2}{8} \sum_{j,k = 1}^n \frac{p_j p_k}{m_j m_k} U_{i j k} + \mathcal O(h^4), \\
\end{aligned}
\label{eq:relateTM}
\end{equation}
}
where $\tilde{H}$ is the midpoint Hamiltonian given by \eqref{Hammid}.  {Note we do not follow Einstein summation convention, but we could restore the convention, for example, by substituting $\partial H/\partial p_i$ for $p_i/m_i $.}
Using this information, we can compute that along the trapezoidal trajectory,{
\begin{equation}
\frac{d}{dt} \tilde{H} = \sum_{i=1}^n \left(\dot{p}_i^\mathrm{T} \frac{\partial \tilde{H}}{\partial p_i} + \dot{q}_i^\mathrm{T} \frac{\partial \tilde{H}}{\partial q_i}\right) = -\frac{h^2}{8} \sum_{i,j,k = 1}^n \frac{p_i p_j p_k}{m_i m_j m_k} U_{i j k} + \mathcal O(h^4).
\label{eq:edrift}
\end{equation}}
This equation describes the energy drift of trapezoidal rule.  The $\mathcal O(h^2)$ term can be integrated with respect to time, and we find that
\begin{equation}
\frac{d}{dt} \tilde{E}_2  = \mathcal O(h^4),
\end{equation}
where
{
\begin{equation}
\label{eq:consesec}
\tilde{E}_2 = H_0 + \frac{h^2}{12} \left(\sum_{i,j = 1}^n U_{i j} \frac{p_i p_j}{m_i m_j} + \sum_{i = 1}^n \frac{1}{m_i}{U_i^2}{} \right):
\end{equation}
}
\noindent the trapezoidal rule has a conserved energy at least to second order.  We will check this numerically in Section \ref{sec:int}. This means a time-symmetric, non-symplectic method can also have a conserved energy at some order, but this fact may not in of itself be useful.  The trapezoidal and midpoint rule, and DKD and KDK leapfrog all have a conserved energy to second order, which, for Hamiltonian \eqref{eq:h0} has form {
\begin{equation}
\tilde{E}_2 = H_0 + h^2 \left(\sum_{i,j = 1}^n  a U_{i j} \frac{p_i p_j}{m_i m_j} + \sum_{i = 1}^nb \frac{1}{m_i}U_i^2 \right),
\label{eq:conse}
\end{equation}
}\noindent and their coefficients $a$ and $b$ are shown in Table \ref{tab:coeff}.  $a$ and $b$ differ from each other for the leapfrog methods because they are partitioned RK methods, as mentioned in Section \ref{sec:RK}.  

\begin{center}
\begin{table}
\caption{The midpoint and trapezoidal rules, and KDK and DKD leapfrogs, have a conserved energy to second order described by \eqref{eq:conse}.  They only differ in the values of the coefficients of $a$ and $b$, whose absolute value is either $1/12$ or $1/24$, and we list them here.}
\centering
\begin{tabular}{| c || c| c|}
	\hline
	 Method & $a$ &$b$  \\ [3ex] \hline
	Midpoint &  $-\frac{1}{24}$  &  $-\frac{1}{24}$\\ 
	Trapezoidal &  $+\frac{1}{12}$  &  $+\frac{1}{12}$\\ 
	KDK Leapfrog & $-\frac{1}{24}$  &  $+\frac{1}{12}$\\ 
	DKD Leapfrog & $+\frac{1}{12}$  &  $-\frac{1}{24}$\\  \hline
	\end{tabular}
	\label{tab:coeff}
\end{table}
\end{center}

We can do better and show that trapezoidal rule has a conserved energy to at least fourth order.  Substituting its MDE \eqref{ftrap} into equations (\ref{Econs1234}), reveals,
\begin{equation}\label{Etrap4}
  \tilde{E}=H+\frac{h^2}{12}(\hat D_{21}H)-\frac{h^4}{720}
      (3\hat D_{40}+6\hat D_{41}-\hat D_{43})H+O(h^6)\ .
\end{equation}
For a conventional Hamiltonian \eqref{eq:broad}, this becomes 
{
\begin{equation}
\begin{aligned}
\label{Etrap5}
  & \tilde{E}=H+\frac{h^2}{12} \left( \sum_{i=1}^n \frac{1}{m_i} U_i^2+\sum_{i,j=1}^n U_{i j} \frac{p_i p_j}{m_i m_j}\right) \\
  &-\frac{h^4}{240} \left[  \sum_{i,j=1}^n \frac{U_i U_j U_{ij}}{m_i m_j}  - \sum_{i,j,k=1}^n \left( \frac{p_ip_k U_{j k} U_{i j}} {m_im_jm_k} + 2\frac{p_ip_j U_k U_{i j k}}{m_i m_j m_k} \right) \right. \nonumber \\
  & + \left.  \frac{1}{3} \sum_{i,j,k,l=1}^n \frac{p_ip_jp_kp_l U_{ijkl}}{m_im_jm_km_l}\right] \nonumber \\
    &+O(h^6) .
    \end{aligned}
\end{equation} }

In fact, we are able to show that the trapezoidal rule conserves an energy function to all orders in $h$, and we can write it down.  Rewrite the trapezoidal rule as a sequence of three RK steps:
\begin{eqnarray}\label{trapfrommid}
  {\bm y}_{-1/2}&=&{\bm y}_0-\frac{1}{2}h{\bm f}({\bm y}_0)
    \nonumber\\
  {\bm y}_{1/2}&=&{\bm y}_{-1/2}+h{\bm f}\left(\frac{1}{2}
    {\bm y}_{-1/2}+\frac{1}{2}{\bm y}_{1/2}\right)
    ={\bm y}_{-1/2}+h{\bm f}({\bm y}_0)\nonumber\\
  {\bm y}_1&=&{\bm y}_{1/2}+\frac{1}{2}h{\bm f}({\bm y}_1)
    ={\bm y}_0+\frac{1}{2}h[{\bm f}({\bm y}_0)+{\bm f}({\bm y}_1)]\ .
\end{eqnarray}
The first step is a backwards explicit Euler step, the second is a symplectic midpoint method, and the third is an implicit Euler step.  Because the implicit midpoint rule has a conserved Hamiltonian (assuming convergence of the series), it is natural to assume that the trapezoidal rule respects an energy function with the same functional form, but with shifted initial conditions.  Indeed, let
\begin{equation}\label{Etrap}
  E_{\rm trap}({\bm y})=\tilde H_{\rm{midpoint}}\left[{\bm y}-\frac{1}{2}h{\bm f(\bm y)}\right].
\end{equation}
Then, we can check $E_{\rm trap}({\bm y_0}) = E_{\rm trap}({\bm y_1})$, which implies that the trapezoidal rule conserves the energy function $E_{\rm trap}({\bm y})$.  If the underlying equations have a quadratic invariant $Q$, we also see the trapezoidal rule has a related invariant,
\begin{equation}\label{Qtrap}
  Q_{\rm trap}({\bm y})=Q\left[{\bm y}-\frac{1}{2}h{\bm f(\bm y)}\right].
\end{equation}

To fourth order, \eqref{Etrap} agrees with equation (\ref{Etrap4}), but it is exact to all orders.   We will derive in {Appendix \ref{sec2}} that there exist time-symmetric methods which are not energy conserving to all orders.  These results are summarized in Table \ref{table:RK}.  This means we can find energy drift with a symmetric integrator with fixed time step{---} this result has already been discussed by \cite{FHP04} and others.

\subsubsection{An example: the simple harmonic oscillator} \label{sec:SHO}
We derive the conserved energy of the trapezoidal rule for the simple harmonic oscillator (SHO).  The Hamiltonian for the SHO is
\begin{equation}
H(q,p) = \frac{1}{2} \left(q^2 + p^2 \right).
\end{equation}
For this Hamiltonian, the trapezoidal rule becomes explicit, since the coordinate derivatives are linear in coordinates.  Also, in this case, the implicit midpoint rule gives an identical rule.  The rules say{
\begin{eqnarray}\label{eq:SHO1}
  {q}_1&=&{ q}_0+ \frac{h}{2}\left(p_0 + p_1\right) \qquad \text{and} \nonumber\\ 
  { p}_1&=&{ p}_0- \frac{h}{2} \left(q_0 + q_1 \right).
\end{eqnarray}
When solved for $q_1$ and $p_1$, they say
\begin{eqnarray}\label{midSHO}
  {q}_1&=&a{ q}_0+ b{ p}_0 \qquad \text{and} \nonumber\\ 
  { p}_1&=&a{ p}_0- b{ q}_0,
\end{eqnarray}}
 where
 \begin{equation}
 a = \left(\frac{1-\delta}{1+\delta}\right), \qquad b = \frac{h}{(1+\delta)}, \qquad \text{and} \qquad \delta = \frac{h^2}{4}.
 \end{equation}
\eqref{midSHO} is also the exact trajectory after time $h$ for a Hamiltonian
\begin{equation} \label{eq:modH}
\tilde{H} = A H,
\end{equation}
so long as
\begin{equation}
\label{eq:cond}
\begin{aligned}
\cos (A h) &= a, \qquad \text{and} \\
\sin (A h) &= b,
\end{aligned}
\end{equation}
implying
\begin{equation}
A = \frac{1}{h} \tan^{-1} \left(\frac{h}{1 - \delta} \right).
\end{equation}
$0< A < 1$ for $0 < h < 2$, so the numerical value of the modified Hamiltonian is smaller than $H$.  When $h \ge 2$, an $A$ satisfying \eqref{eq:cond} does not exist, so the governing equations are no longer Hamiltonian.  Thus, the trapezoidal and implicit midpoint rules' MDEs are governed by \eqref{eq:modH}.  This implies they exactly conserve the energy of the SHO, as one can verify numerically.
 
 For a general Hamiltonian (e.g. the H\'{e}non{--}Heiles problem), these simple exact results no longer hold. However, for a time-independent Hamiltonian, symplectic methods always have a conserved energy, and so do conjugate methods like the trapezoidal rule.

\subsection{Modified differential equation with adaptive time steps} \label{sec:modad}
In previous sections and the Appendix, we discuss integrators with fixed step-sizes, but for fixed step-sizes, there already exist excellent symplectic integrators in astrophysics, starting with leapfrog.  Time-symmetric integrators are popular in astrophysics due to their ability to accommodate adaptive stepping.  An exactly time-symmetric integrator was proposed by \cite{HMM95}, and approximately time-symmetric integrators have been developed by \cite{pelu12} and \cite{K98}.
We focus on the proposal by \cite{HMM95}, because it is exactly time-symmetric, under certain conditions we describe.  In conjunction with leapfrog, they propose to write the time step as an implicit equation,
\begin{equation}
h = \frac{\epsilon}{2} \left[\sigma(\v{y}_0) + \sigma(\v{y}_1) \right].
\label{eq:adcrit}
\end{equation}
$\sigma(\v{y})$ is a function that we can specify using a priori knowledge about the solution trajectory (e.g., the relevant timescales) or even without this knowledge \citep{S95}.  An implicit step criterion can be used with an implicit one-step method, like the trapezoidal rule, not necessarily resulting in more iterations when solving the update equations. \eqref{eq:adcrit} can be written as an explicit infinite series in $\epsilon$, 
\begin{equation}
h = \epsilon s(\v{y}_0,\epsilon) = \epsilon s_0 (\v{y}_0) + \epsilon^2 s_1 (\v{y}_0) + \hdots
\label{eq:ad}
\end{equation}
The direction of time is now determined by the sign of $\epsilon$.  Of course, the ${s}_i$ depend on the method.  Letting $\sigma \equiv \sigma(\v{y}_0)$, for trapezoidal rule,
{
\begin{equation}
\begin{aligned}
s_0 &= \sigma, \\
s_1 &= \frac{1}{2} \sum_{i = 1}^{2n} \sigma f_i \partial_i \sigma, \\
s_2 &= \frac{1}{4} \left[   \sigma \sum_{i= 1}^{2n} (f_i \partial_i \sigma)^2 + \sigma^2  \sum_{i,j=1}^{2n}\left( (f_j \partial_j f_i) \partial_i \sigma + f_j f_i \partial_j \partial_i \sigma \right) \right], \\
& \vdots,
\end{aligned}
\end{equation} }
\noindent {where $2n$ is the phase space dimension.  $f_i$ is the $i$th component of $\v{f}$ in eq. \eqref{sysode} and $\partial_i \equiv \partial / \partial y_i$}.  Symmetry requires 
\begin{equation}
\label{eq:symm}
s(\v{y}_1,-\epsilon) = s(\v{y}_0, \epsilon).
\end{equation}
For criterion \eqref{eq:adcrit}, this requirement is automatically satisfied.  $\rho${-}reversibility would require $s(\bm{\rho} \v{y}_1,\epsilon) = s(\v{y}_0,\epsilon)$.  For steps \eqref{eq:adcrit} this means \citep[Section VIII.3]{hair06}
\begin{equation}
\sigma(\bm{\rho} \v{y}) = \sigma(\v{y}).
\label{eq:sigma}
\end{equation}
This condition is easy to satisfy, but it is not always satisfied.  \cite{HMM95} {were} interested in the $N$-body problem and  proposed a $\sigma$ that is the minimum of the close encounter and free fall times.  This satisfies \eqref{eq:sigma} if we are taking the absolute values of relative velocities.  We will explore what happens when eq. \eqref{eq:sigma} is not obeyed in Section \ref{sec:numdem}.   The equations \eqref{eq:adcrit}, \eqref{eq:symm}, and \eqref{eq:sigma} apply whether the underlying method is a{n} RK method, like trapezoidal rule, or a partitioned Runge{--}Kutta method, like leapfrog.  But the underlying method must be time-symmetric for either \eqref{eq:symm} or \eqref{eq:sigma} to hold.

{The stepping rule \eqref{eq:adcrit} is implicit, which is more cumbersome to analyze than an explicit rule.  However, we use an implicit criterion for the following reasons:  
\begin{itemize}
\item{The trapezoidal rule is already implicit, so choice \eqref{eq:adcrit} does not necessarily add more computational work.}
\item{\eqref{eq:adcrit} has already been used by \cite{HMM95} and others.}
\item{\cite{D17} studies explicit stepping criteria with step sizes that can only take certain values.  The discreteness of the step sizes breaks the time-symmetry and reversibility symmetries.  We want to construct exactly symmetric and reversible methods for our tests to be able to conclude that errors are not due to breaks in these symmetries.}
\item{The explicit stepping criteria in \cite[Sections 4 and 5]{D17}, even in the continuous, non-discrete case, risk becoming unsynchronized with $\sigma$, leading to stepping of questionable efficiency and accuracy.  This stepping can be regarded as a multistep method.}
\end{itemize}
}

We can construct the MDE with adaptive time steps, following the procedure of Section \ref{sec:mde} and using the form \eqref{eq:ad}, so that the series are now written in terms of $\epsilon$.  Now, instead of eq. \eqref{eq:relateTM}, we have
{
\begin{equation}
\begin{aligned}
q_i^\mathrm{T} &=  q_i^\mathrm{M} - \frac{1}{12} (\epsilon s_0)^3 \sum_{j = 1}^n \frac{p_j}{m_i m_j} U_{i j} +  \mathcal O(\epsilon^4), \\
p_i^\mathrm{T} &= p_i^\mathrm{M} - \frac{1}{12} (\epsilon s_0)^3  \left( \sum_{j,k = 1}^n \frac{p_j p_k}{m_j m_k} U_{i j k} - \sum_{j=1}^n U_{i j} U_j \right)+  \mathcal O(\epsilon^4), \\
\dot{q}_i^\mathrm{T} &=  \frac{\partial {H_0} }{\partial {p_i}}  -  \frac{1}{12} (\epsilon s_0)^2 \sum_{j = 1}^n \frac{p_j}{m_i m_j} U_{i j} \mathcal + O(\epsilon^3), \\
\dot{p}_i^\mathrm{T} &= - \frac{\partial {H_0}}{\partial {q_i}} - \frac{1}{12} (\epsilon s_0)^2 \left(   \sum_{j,k = 1}^n \frac{p_j p_k}{m_j m_k} U_{i j k} - \sum_{j=1}^n U_{i j} U_j \right) + \mathcal O(\epsilon^3). \\
\end{aligned}
\end{equation} }
As in eq. \eqref{eq:edrift}, we can calculate the energy drift along the trapezoidal orbit as
{
\begin{equation}
\frac{d}{dt} H_0 =-\frac{1}{12} \epsilon^2 \left[\sigma(\v{q},\v{p})\right]^2 \sum_{i,j,k = 1}^n U_{i j k} \frac{p_i p_j p_k}{m_i m_j m_k} + \mathcal O(\epsilon^3).
\label{eq:edriftad}
\end{equation} }
If we let $\sigma = 1$ and $\epsilon = h$, this expression is just \eqref{eq:edrift}.  This shows that the energy drift is a function of the problem and the choice of $\sigma$.  In general, this $\epsilon^2$ term cannot be integrated in terms of elementary functions.  There exist reversible $\sigma$ (cf. eq. \ref{eq:sigma}) which lead to secular drift and irreversible $\sigma$ which lead to no energy drift, as we will show in Section \ref{sec:numdem}.
\eqref{eq:edriftad} holds for any {$h(\epsilon,\v{y}_0)$}, not just \eqref{eq:adcrit}, so long as to lowest order in $\epsilon$, {$h = \epsilon \sigma(\v{y}_0)$}.  For example, \eqref{eq:edriftad} applies to the geometric mean time step,
{
\begin{equation}
\label{eq:geom}
h = \epsilon \left[\sigma(\v{y}_0) \sigma(\v{y}_1) \right]^{1/2}.
\end{equation}
}

\section{Numerical Demonstration}
\label{sec:numdem}
In this section we apply the error analysis of Section \ref{sec:trap} to see that energy conservation is violated in a number of situations, even when a method is symmetric and reversible.  For the tests, we consider the pendulum solution and H\'{e}non{--}Heiles orbits.  
\subsection{The modified pendulum}
\label{sec:int}
In the following, we consider the trapezoidal rule \eqref{eq:trap} along with the adaptive step criteria \eqref{eq:adcrit}.   Consider the simple pendulum, with a modified potential:
\begin{equation}
\label{eq:modpend}
H = \frac{p^2}{2} - \cos{q} + \frac{1}{5} \sin (2 q).
\end{equation}
This modified pendulum was considered by \cite{FHP04}.  The reason for choosing this potential with $1/5 \sin(2 q) $ will become apparent below.  This Hamiltonian is $\rho${-}reversible: $\frac{\partial H(q,-p)}{\partial p} = - \frac{\partial H(q,p)}{\partial p}$, and $\frac{\partial H(q,-p)}{\partial q} = + \frac{\partial H(q,p)}{\partial q}$.   

\begin{figure}
	\begin{center}
	\includegraphics[width=1\columnwidth]{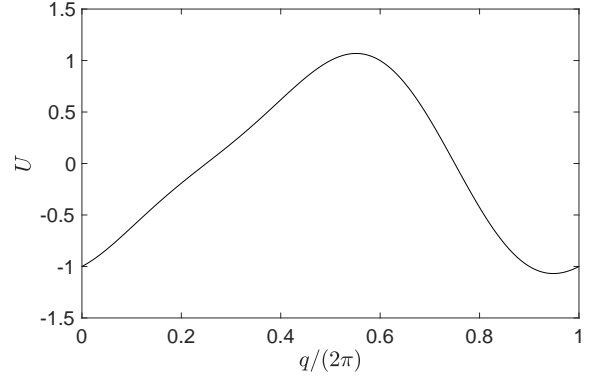}
	\end{center}
	\vspace*{-3mm}
	\caption{Potential as a function of the periodic range of $q$ for the modified pendulum Hamiltonian \eqref{eq:modpend}.  There is no symmetry in the potential and it has a minimum of $U \approx -1.069$.}
	\label{fig:pendpot}
\end{figure}

\begin{figure}
	\begin{center}
	\includegraphics[width=1\columnwidth]{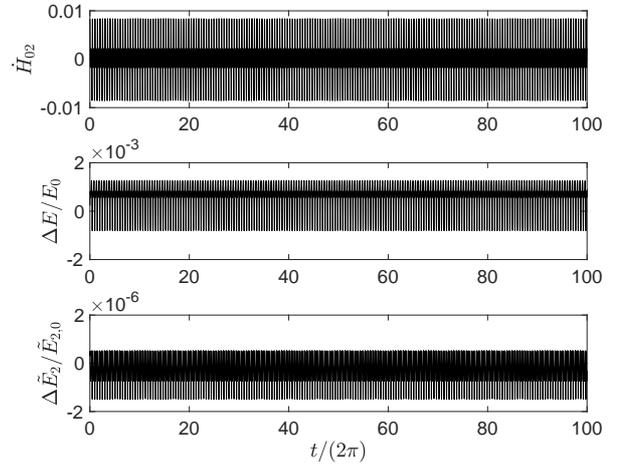}
	\end{center}
	\vspace*{-3mm}
	\caption{The evolution of some phase space quantities as a function of time when we integrate the modified pendulum with the symmetric non-symplectic trapezoidal rule.  We use a constant $h \approx 0.63$ and initial conditions $p = 2.5$ and $q = 0$.  The top panel gives the change in energy as given by \eqref{eq:edriftad}.  The middle panel gives the energy error, and the bottom panel gives the error of the conserved second order energy.  No energy drift is observed in the middle panel, in agreement with the top panel.  The second order energy is conserved better than the energy, as expected.}
	\label{fig:modpenderror}
\end{figure}

The potential is not symmetric in $q$ over the periodic range of $q$, as seen in Fig. \ref{fig:pendpot}.  The minimum is $U \approx -1.069$ and occurs at $q \approx 5.959$.  
First, we choose $\sigma(\v{y}) = 1$, so that the step is constant.  Because this $\sigma$ satisfies \eqref{eq:sigma}, the integrator is reversible.  We choose $h = \epsilon = 2 \pi/100 \approx 0.63$: there are roughly 100 steps per period.  We let $t_{\mathrm{final}} = 100$.  As initial conditions, we choose $p = 2.5$ and $q = 0$ so that $H = 2.125$.  So for the exact solution (and in all our numerical solutions), the sign of the momentum does not change {(the orbit is circulating)}.  In Fig. \ref{fig:modpenderror}, we show that the change in various phase space quantities in time mimics the behavior of a symplectic integrator.
{$\dot{H}_{02}$} is given by the second order $\epsilon^2$ term of eq. \eqref{eq:edriftad}:
{
\begin{equation}
\dot{H}_{02} = -\frac{1}{12} \epsilon^2 \left[\sigma(\v{q},\v{p})\right]^2 \sum_{i,j,k = 1}^n U_{i j k} \frac{p_i p_j p_k}{m_i m_j m_k},
\end{equation}
}
\noindent {which oscillates} symmetrically around 0.  We also show the energy and $\tilde{E}_2$ error, from \eqref{eq:conse}.  $E_0$ and $\tilde{E}_{2,0}$ are the initial energy and $\tilde{E}_2$ values, respectively.  $\tilde{E}_2$ is conserved better than $E$, supporting the finding that a conserved energy exists.  We checked that, for a fixed integration time, $\Delta E/E \propto h^2$ and $\Delta \tilde{E}_2/\tilde{E}_2 \propto h^4 $.

\begin{figure}
	\begin{center}
	\includegraphics[width=1\columnwidth]{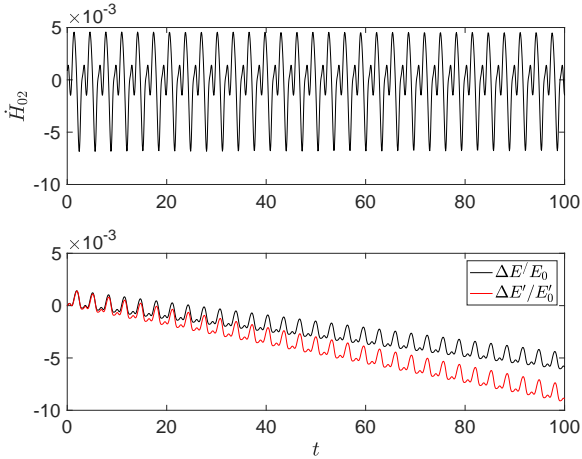}
	\end{center}
	\vspace*{-3mm}
	\caption{{Same as} Figure \ref{fig:modpenderror}, but now with a reversible, time-symmetric step size strategy, $\sigma(\v{y}) = U + 1.5$.  $\epsilon \approx 0.039$.  We observe a linear energy drift, which roughly agrees with the prediction from the top panel, given by the red curve in the lower panel.  This integration is fully symmetric and reversible, yet shows a linear energy error drift.}.
	\label{fig:modpendad}
\end{figure}

We next choose an adaptive step strategy.  We choose $\sigma(\v{y}) = U + 1.5$, so that $\sigma(\v{y}) > 0$.  This choice is both reversible and time-symmetric, according to the discussion above \eqref{eq:sigma}.  We checked if we integrate forwards, change the sign of $p$, and integrate the same number of steps backwards, we recover the initial conditions {up to roundoff error}.  We choose $\epsilon = 2 \pi/(100 \times 1.63)$ so that the average time step is approximately still the same as the previous test.  {The initial phase space coordinates remain the same in this test: $p = 2.5$ and $q = 0$.}  We initialize the integration with guess for the initial step, $h_0 = \epsilon \sigma(\v{y}_0)$, and thereafter we use the previous time step as the initial guess.  We integrate for the same total time as Fig. \ref{fig:modpenderror}, and show the results in Fig. \ref{fig:modpendad}.  
{$\dot{H}_{02}$} now is not symmetric around 0.  This leads to a linear drift in energy error as seen on the bottom subplot.  If {$\dot{H}_{02,i}$} is the value of $\dot{H}_0$ at time step {number} $i$, and $h_i$ is the value of the time step, define
{
\begin{equation}
E^\prime_m = E_0 + \sum_{i=1}^m h_i \dot{H}_{02,i},
\end{equation}
}
\noindent which we expect to be close to $E_m$, the energy at step $m$.  We see in Fig. \ref{fig:modpendad} this is the case at small time, but the approximation breaks down for larger time.  As we decrease $\epsilon$, the difference between the two curves becomes undetectable on the same type of plot.  We checked the slope of the $\Delta E/E$ curve scales as $\epsilon^2${, confirming the {energy} error is $\mathcal O(t\epsilon^2)$}.  All other $\sigma(\v{y})$ we tested, reversible and irreversible, gave a linear drift in energy for this problem.  For the geometric mean time step \eqref{eq:geom}, the errors are similar, as expected.  The final energy error at {$t = 100 \times 2 \pi$} changes by less than $1\%$. 

{
We integrate \eqref{eq:modpend} with one-step symplectic methods of different properties.  We use stepsize  $h = 2 \pi/100$ and initial conditions $p = 2.5$ and $q = 0$ (as in Section \ref{sec:int}).  The methods are described in Table \ref{tab:modpend}, where we write the method's name, order, Runge--Kutta classification, and whether the method is time-symmetric.  None of the methods yield energy drift.  
\begin{table}
\caption{{Description of symplectic integrators used in tests of the modified pendulum Hamiltonian.}}
\centering
\begin{tabular}{lccc}
\hline\hline
Method & Symmetric? & Order & Classification\\
\hline
Leapfrog DKD and KDK & Yes & 2 & Partitioned Runge--Kutta \\
\hline
Gauss-Legendre & Yes & 4 & Runge--Kutta \\
\hline
Symplectic Euler & No & 1 & Runge--Kutta \\
\hline
\label{tab:modpend}
\end{tabular}
\end{table}

By contrast, it has been reported that Lobatto IIIA and IIIB, two fourth order symmetric and reversible, but non-symplectic methods show energy drift when used with fixed step size.  The tests were performed in \cite{FHP04}, using $h = 0.16$ and $t_{\mathrm{max}} = 1600$.  \cite{FHP04} differs from our work in that it considered only fixed time-steps.  We confirmed the leading order error for Lobatto IIIB is $\mathcal O(t h^4)$ while the leading order error for Lobatto IIIA is $\mathcal O(h^4)$ and has no error drift (at leading order), except for roundoff error contributions.  Lobatto IIIA and IIIB have the same symmetries as the trapezoidal rule with adaptive symmetric and reversible steps: time-symmetry and time-reversibility.

The simplified Takahashi--Imada method is symmetric and volume preserving.  This means it conserves one Poincar\'{e} invariant, which does not guarantee symplecticity.  It was reported in \cite{HMS09} that this method gives secular drift in a function close to the energy for Hamiltonian \eqref{eq:modpend}.  They used $h = 0.2$ and $t_{\mathrm{max}} \approx 1900$.  In our own tests, we found the method gives energy drift.  Volume preservation is equivalent to symplecticity for one-degree-of-freedom problems such as Hamiltonian \eqref{eq:modpend}, so this would appear to be an example of a symplectic integrator giving energy drift.  But this integrator is generally non-symplectic.

The only example we found of a reversible, symmetric, and non-symplectic method conserving energy for this problem is studied in Fig. \ref{fig:modpenderror}.  We have not tested symplectic integrators with adaptive steps because symplecticity is not conserved and this advantage of the method is lost.

When the orbits of these initial conditions are computed with symplectic integrators, we have observed that different steps along the orbit produce increases or decreases in energy.  The net increase is zero.  For the orbit of Fig. \ref{fig:modpendad}, the net increase is negative over an orbit, leading to energy drift.}

For the unmodified pendulum,
\begin{equation}
H = \frac{p^2}{2} - \cos(q),
\end{equation}
{consider the same initial conditions given above.  The sign of $p$ is still invariant in all tests.}  {B}oth reversible and irreversible $\sigma$, such as $\sigma(\v{y}) = 1.5 -\cos(q)$ and $\sigma(\v{y}) = ap + b$ with $a$ and $b$ constants, give no drift in energy error.  However, we again get drift if we let $\sigma(\v{y})$ be an asymmetric function of $q$, such as the {$\sigma(\v{y}) = 1.5 - \cos(q) + 1/5 \sin(2 q)$} we used above, even though this $\sigma$ is reversible.  These experiments show that time reversibility and energy conservation are independent concepts.  {To summarize, for circulating pendulum orbits, all time-symmetric methods, except the fixed time trapezoidal rule, gave undesirable error behavior.  This exception may be related to the fact it is a conjugate symplectic method (see Section \ref{sec:rel}).  All tested symplectic methods except the simplified Takahashi-Imada method yielded desirable energy conservation.  The simplified Takahashi-Imada method is generally non-symplectic.}  

\subsection{H\'{e}non{--}Heiles orbits}
\begin{figure}
	\begin{center}
	\includegraphics[width=1\columnwidth]{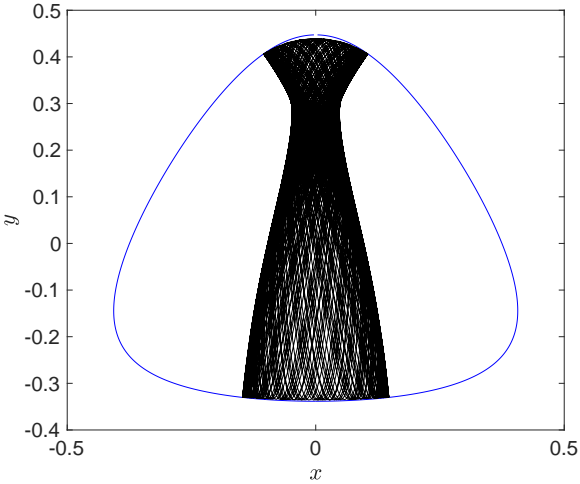}
	\end{center}
	\vspace*{-3mm}
	\caption{A regular box H\'{e}non{--}Heiles orbit.  The initial conditions are $x = 0$, $y = 0.2$, $p_y = 0.3$, and $H = 0.07019755555555$.  We also plot the bounding equipotential.  The integration is run until $t_{\mathrm{max}} = 100 \times 2 \pi$ with a constant step-size trapezoidal method.}
	\label{fig:hhtraj}
\end{figure}

\begin{figure}
	\begin{center}
	\includegraphics[width=1\columnwidth]{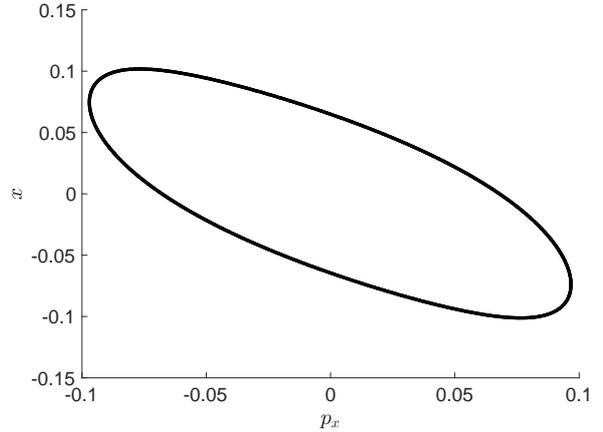}
	\end{center}
	\vspace*{-3mm}
	\caption{Surface of section plot for the orbit of Fig. \ref{fig:hhtraj}.  A point is plotted everytime $y = 0$ is crossed with $p_y > 0$.  The symmetry in $p_x$ indicates a box orbit, and the closed curve indicates a regular orbit far from resonance.  The surface of section is computed up to time $t = 10^5$ with a high accuracy Runge{--}Kutta method.}.
	\label{fig:chaotsurf}
\end{figure}

The H\'{e}non{--}Heiles problem Hamiltonian \citep{HH64} is a two-degree-of-freedom problem{---}a simplified model of a galactic potential.  The Hamiltonian is 
\begin{equation}
\label{eq:hh}
H = \frac{1}{2} (p_x^2 + p_y^2) + U(x,y)
\end{equation}
with {$U(x,y) = \frac{1}{2} (x^2 + y^2) + x^2y - \frac{y^3}{3}$}.  This Hamiltonian allows both chaotic and regular trajectories and is $\rho$-reversible.  We consider a regular orbit with initial conditions $x = 0$, $y = 0.2$, $p_y = 0.3$, {$p_x = 0.125413095187199$}, and $H = 0.07019755555555$ (although we only need to keep three significant figures to get the same qualitative results).  Using $\sigma = 1$, and $\epsilon = h = 0.1$, we show the trajectory in Fig. \ref{fig:hhtraj} with $t_{\mathrm{max}} = 100 \times 2 \pi$.  Also plotted is the bounding equipotential curve, $U = H$.   
The trajectory does not span the entire allowed area, which tells us it is a regular orbit.  We can verify this in a surface of section plot.  In Fig. \ref{fig:chaotsurf}, we plot a point in the $x${--}$p_x$ plane every time $y = 0$ is crossed with $p_y > 0$, up to time $t = 10^5$.  We use a fifth and sixth order pair of Runge{--}Kutta methods{, in what's known as Verner's embedded Runge--Kutta method \citep{Verner78},} for this plot.  This is an adaptive step method: two methods allow an estimate of the local truncation error which is then used to determine a step size.  The final energy error is $\approx 3.2 \times 10^{-14}$.  This orbit is a box orbit: the sign (and magnitude) of the angular momentum oscillates.  If we change the sign of the initial momenta, the trajectory is confined to the same bounding curve, which means the second isolating integral besides the energy (for these initial conditions) does not depend on the sign of the momenta.  This is a consequence of the $\rho$-reversibility of the equations due to \eqref{eq:hh}.  We checked this by running the trajectory with a sign change in the initial momenta and checking that the minimum and maximum $x$ of the trajectory is the same to 15 significant figures.  

\begin{figure}
	\begin{center}
	\includegraphics[width=1\columnwidth]{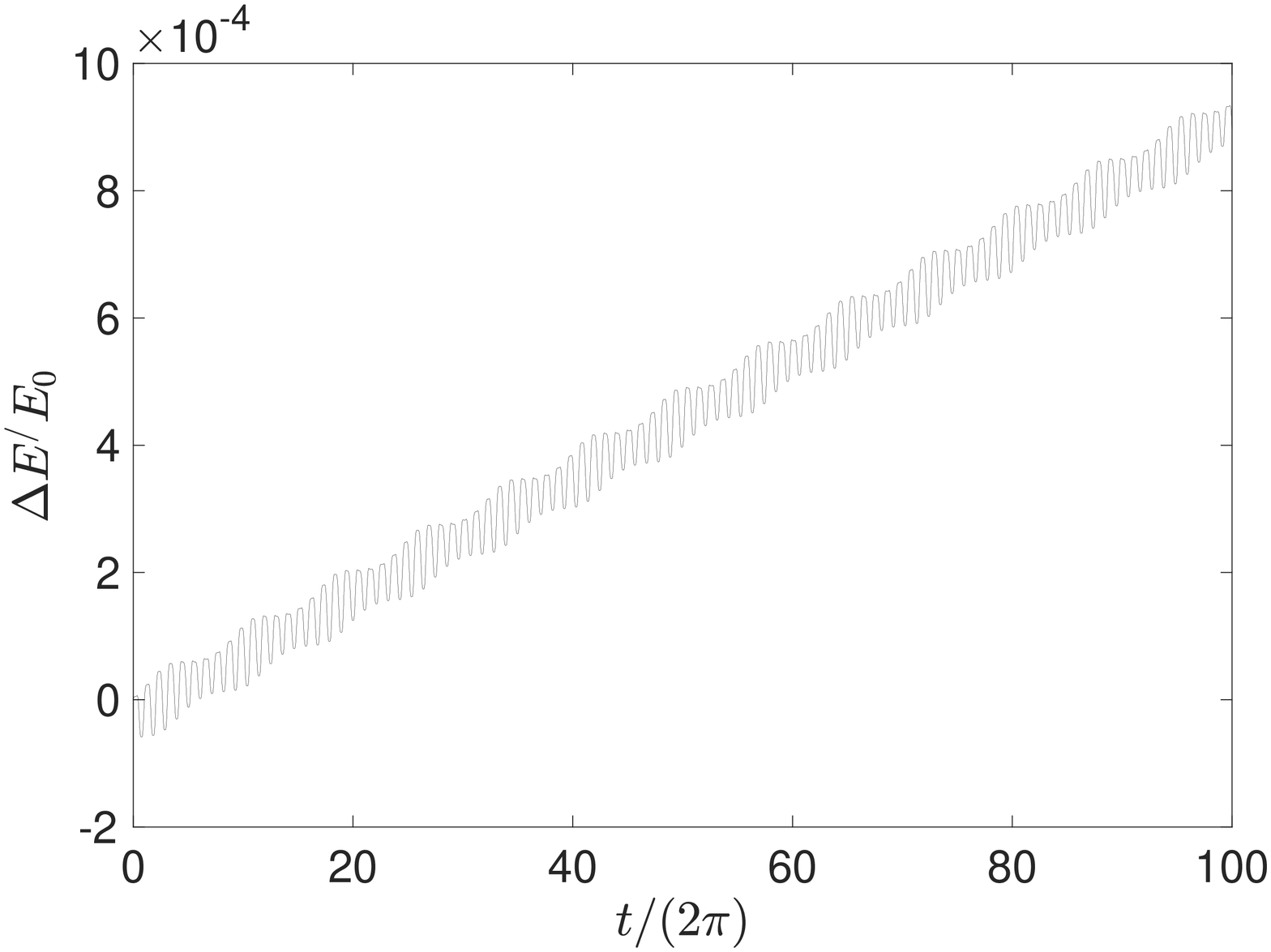}
	\end{center}
	\vspace*{-3mm}
	\caption{Energy error over time for the orbit of Fig's. \ref{fig:chaotsurf} and \ref{fig:hhtraj}.  We run a trapezoidal rule with step selector $\sigma = 10^{-3} p_y + 0.01$ and $\epsilon = 2.5$.  This integrator is time-symmetric, but not reversible, and shows a linear drift in energy.  The energy drift is predicted accurately by \eqref{eq:edriftad}.}
	\label{fig:hh}
\end{figure}

For this Hamiltonian, for eq. \eqref{eq:edriftad}, we have
\begin{equation}
\frac{d}{dt}H_0 = -\frac{1}{12} \epsilon^2 \sigma^2(\v{y}) \left[2p_y (3 p_x^2 - 2 p_y^2)  \right].
\end{equation}
Note the asymmetry in $p_y$ and $p_x$ due to the potential.  For the orbit of {Figure \ref{fig:hhtraj}}, $\overline{y} \approx 0.07 $.  So the centroid is non-zero for this orbit.  We choose $\sigma(\v{y}) = a p_y + b$, where $a = 10^{-3}$ and $b = 10^{-2}$, to ensure $\sigma(\v{y}) > 0$.  This $\sigma$ is irreversible (below we will explore other $\sigma$, reversible and irreversible).  We let $\epsilon = 2.5$ and $t_{\mathrm{max}} = 628$.  We can estimate the typical time step by using
\begin{equation}
\overline{h} \approx \epsilon(0.01 + 0.001 \overline p_y).
\end{equation} 
We measure experimentally a time weighted average of $p_y$ of $\overline{p_y} \approx -5 \times 10^{-4}$, which gives $\overline{h} \approx 0.025$, in agreement with experiment.  In Fig. \ref{fig:hh}, we show the error in energy over time{---}it has a linear drift with slope about $1.4 \times 10^{-6}$.  We checked, by varying $\epsilon$, that the slope scales with $\epsilon^2$.
If we plot the error in $E^\prime$ on the same plot, it is nearly indistinguishable from the error in $E$.  If $\delta = \Delta E/E - \Delta \tilde{E}/\tilde{E}$, $\overline{\delta} = 4.3 \times 10^{-7}$.  For $\sigma(\v{y}) = a p_y^n + b$, there will only be a drift for $n$ odd.  When $n$ is even, the integrator is again reversible and drift is eliminated for this problem.  

\begin{figure}
	\begin{center}
	\includegraphics[width=1\columnwidth]{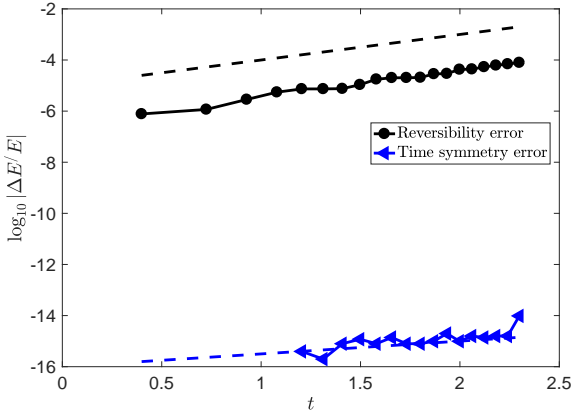}
	\end{center}
	\vspace*{-3mm}
	\caption{The time-symmetry and time-reversibility error for the orbit of Fig. \ref{fig:hh}.  To compute the time symmetry error, we integrate forward for $t$, switch the sign of $\epsilon$ and integrate forward the same number of steps.  For the reversibility error, we change the sign of {$\v{p}$} instead.  The former error indicates the integrator is time-symmetric and the error grows as $t^{1/2}$ as expected.  The latter error grows as $t$ and shows reversibility is broken.}
	\label{fig:revers}
\end{figure}

We show this integrator is time-symmetric but not reversible in Fig. \ref{fig:revers}.  Here, after choosing an $\epsilon$, we run forwards for some given time.  Then, we switch the sign of $\epsilon$ and run forwards the same number of steps.  We repeat the experiment for various $\epsilon$, and plot $\epsilon$ vs the energy error.  For the second curve, labelled after running forwards, we change the sign of {$\v{p}$} instead of $\epsilon$.  The errors of the first experiment are small, given by roundoff error, and indicated as the ``Time symmetry'' error in Fig. \ref{fig:revers}.  We measure the error energy of this operation.  In the second case, we change the sign of {$\v{p}$} and run forward the same number of steps.  The first experiment gives a small error, at the level of roundoff.  The dashed blue line shows a slope $t^{1/2}$, which is the expected error growth, based on Brouwer's Law \citep{B37}.  Even though the Brouwer's Law analysis only works for fixed time steps, in a run with $t \approx 39.77$, the standard deviation in the time step lengths is $6.3 \times 10^{-4}$, so this approximation is valid.  The reversibility error shows that this integrator is not reversible.  The dashed black line indicates a slope of $t^1$, as expected from the linear drift of Fig. \ref{fig:hh} (there is a similar linear drift in Fig. \ref{fig:hh} if we initialize with reversed {$\v{p}$}).  It is also possible to develop an integrator that is reversible, but not time-symmetric.  For example, modify \eqref{eq:adcrit} to,
\begin{equation}
\label{eq:unsym}
h = \frac{\epsilon^2}{2} \left[ \sigma(\v{y}_0) + \sigma(\v{y}_1) \right],
\end{equation}
with {$\sigma(\v{y})$} reversible.  According to \eqref{eq:symm}, this breaks time symmetry{, but one might argue this choice is not sensible since it does not allow changing the sign of $h$ with $\epsilon$}.  But this time-symmetry break does not cause any new linear error drift.  {To get} the analogue of \eqref{eq:edriftad} for step \eqref{eq:unsym}: replace $\epsilon^2$ with $\epsilon^4$ in \eqref{eq:edriftad}. 

{Note we have not proved that all bound H\'{e}non{--}Heiles orbits computed with a reversible integrator show no energy drift.  It is possible that for some initial conditions, the solution of the MDEs is a bound orbit such that the energy has a secular increase or decrease, as was the case of Hamiltonian \eqref{eq:modpend}.   All tested unbound orbits resulting from reversible and symmetric trapezoidal methods as well as from symplectic methods give secular energy change.}

\begin{figure*}
	\begin{center}
	\includegraphics[width=1.5\columnwidth]{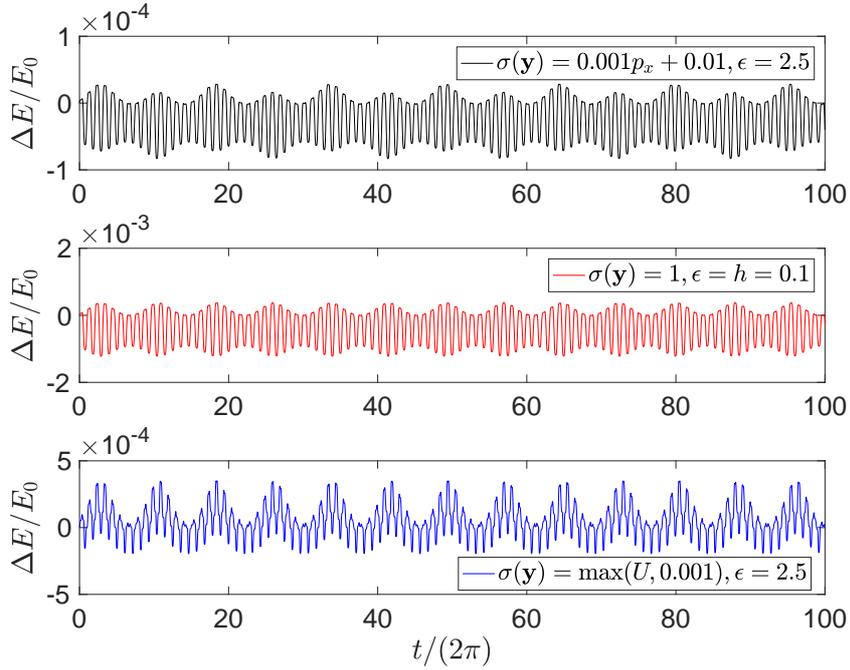}
	\end{center}
	\vspace*{-3mm}
	\caption{A repeat of the experiment of Fig. \ref{fig:hh}.  The initial conditions are the same, but we vary $\epsilon$ and the step criteria $\sigma$.  In no case do we observe energy drift.  All panels show a time-symmetric integration, but only the second and third panel show reversible integration.  This example shows that irreversibility does not imply energy drift.}
	\label{fig:hhdiffstrat}
\end{figure*}

Now we repeat the experiment for different choices of $\sigma(\v{y})$.  In Fig. \ref{fig:hhdiffstrat}, we plot the error in energy vs. time, analogously to Fig. \ref{fig:hh}, but for these different choices of $\sigma(\v{y})$, one irreversible and two reversible.  No linear energy drift is observed in any case.  We also used an explicit second order Runge{--}Kutta method to integrate the orbit, the explicit midpoint rule.  In the notation of Section \ref{sec:RK}, this method has $c = 1$, $b_1=0$, $b_2 = 1$, $a_{2 1} = 1/2$, and $s = 2$.  Using $h=0.1$, we get linear drift in energy error, as expected from standard numerical analysis.  {These experiments demonstrate that reversibility or symplecticity of a method is not a requirement for energy conservation.  As supported in the experiments and shown in Appendix \ref{sec2}, time-symmetry and time-reversibility are properties independent to energy conservation.}

\begin{figure}
	\begin{center}
	\includegraphics[width=1\columnwidth]{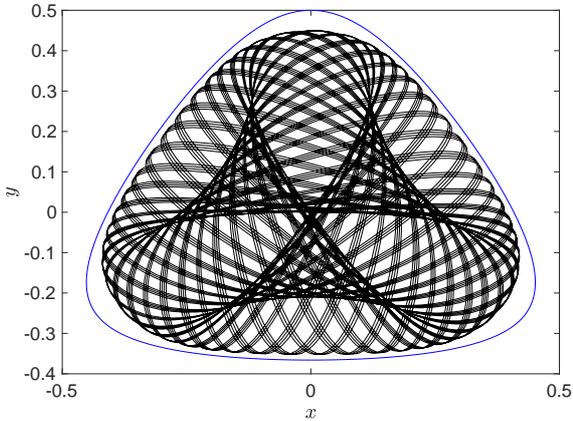}
	\end{center}
	\vspace*{-3mm}
	\caption{A regular loop orbit of the H\'{e}non{--}Heiles problem and the bouding equipotential curve.   The initial conditions are $p_x = \sqrt{1/6}$, $p_y = x = y = 0$ (and $H = 1/12$).  The orbit is plotted until $t_{\mathrm{max}} = 100 \times 2 \pi$.}
	\label{fig:trajhhreg}
\end{figure}

\begin{figure}
	\begin{center}
	\includegraphics[width=1\columnwidth]{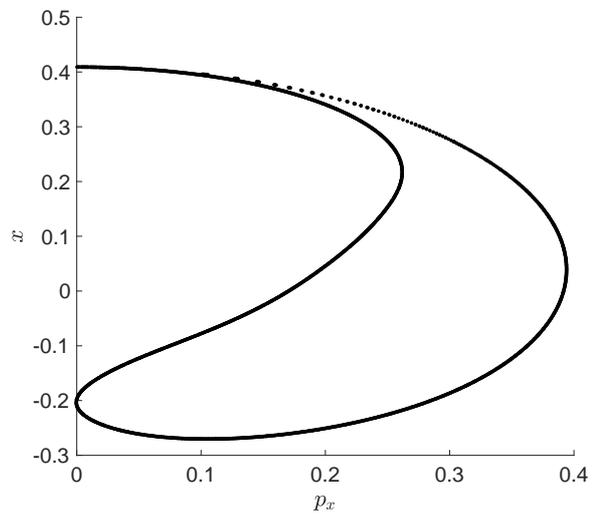}
	\end{center}
	\vspace*{-3mm}
	\caption{Surface of section plot for the orbit of Fig. \ref{fig:trajhhreg}.  The plot is constructed in a similar way to Fig. \ref{fig:chaotsurf}.  The discrete islands indicate the orbit is regular and near a resonance, and the asymmetry in $p_x$ indicates a loop orbit.}.
	\label{fig:sosreg}
\end{figure}

We tested another regular orbit with $H = 1/12$ and initial conditions $p_x = \sqrt{1/6}$, $p_y = x = y = 0$.  This time, the centroid {(mean position in the $x$--$y$ plane)} is at 0.  We plot the trajectory and equipotential curve, $U = 1/12$ in Fig. \ref{fig:trajhhreg}, using $\sigma = 1$, $\epsilon = 0.1$, and $t_{\mathrm{max}} = 200 \pi$.  The trajectory plot indicates this orbit is not far from a periodic resonance orbit, at the boundary between regular and chaotic orbits.  We plot the surface of section in Fig. \ref{fig:sosreg}, again using the pair of Runge{--}Kutta methods with adaptive stepping, for $t = 10^5$, which gives a $7.4 \times 10^{-14}$ energy error. This is a loop orbit: the angular momentum is less than 0 for all time.  For this orbit, we did not find any reasonable $\sigma(\v{y})$ that yields drift, whether $\sigma$ is reversible or irreversible.  

All chaotic orbits we tested show drift in energy, whether $\sigma$ is constant or not.  The drift increases as $t^x$ where $0<x \le 1$.  Chaotic orbits have been investigated elsewhere \citep{HMS09,MP04,HMM95}; typically random walk behavior in the error ($\propto t^{1/2}$) is observed.  We did not find any case in which a chaotic orbit gave a long term linear drift in energy error, as in the previous experiments with regular orbits.  

These experiments for H\'{e}non{--}Heiles show that good energy behavior is possible even with an irreversible integrator.  They also show the {range of} appropriate step criterion{s}{ :they depend on the orbit and problem and are not necessarily restrictive}.  {To summarize, box orbit initial conditions mapped with a time-symmetric but irreversible integrator resulted in energy drift.  But other irreversible integrators yielded no energy drift for this box orbit.  A loop orbit did not give error drift for any tested integrator.  All chaotic orbits gave some form of energy drift in all tests.}

\section{Conclusion} \label{sec:conc} 
This work provides the error analysis needed to understand energy errors of symmetric integrators with adaptive steps used in astrophysics.  We show how to study integrators using their modified differential equations (MDEs) and use this machinery to derive the MDEs for the trapezoidal rule with adaptive steps.  The trapezoidal rule is a time-symmetric, but non-symplectic, integrator.  Other authors have used the leapfrog method with adaptive steps; we do not study this because there is no advantage to using leapfrog as far as error properties are concerned, as discussed in Section \ref{sec:RK} .We find that the trapezoidal rule, with adaptive steps, does not conserve the energy well for some problems, and we use the MDEs to explain this result.  {We cannot make broad statements about the energy conservation of time-symmetric methods because they have different MDEs from each other.  The error of a symmetric integrator depends on the integrator, the differential equation, and the initial values.  But there is no reason to think other methods will not suffer from the same shortcomings of the trapezoidal rule.}

We also note that time-symmetry and reversibility are distinct concepts for an integrator{\cite[Section VIII.3]{hair06}}.  Time symmetry means that if we reverse the sign of the time step, we can recover the initial conditions, while reversibility means that if we switch the sign of velocities and integrate forwards, we will recover the initial conditions.  For the $N$-body problem with pairwise forces, which is reversible, there are integrators which are correctly both reversible and symmetric, neither reversible nor symmetric, or only symmetric or reversible{; we study several of these combinations and the errors they lead to.  For example, we find reversibility does not have to be a requirement for conserving the energy of a H\'{e}non{--}Heiles orbit}.    

We conclude that while time-symmetric integration has often been observed to yield small errors over long time-scales when used for the $N$-body problem, it is not always the case that a time-symmetric integration will work successfully.  We suggest that caution be used when deciding to use a time-symmetric method, and that preference should still be given to symplectic integrators.  In general, time-symmetric methods are not guaranteed to conserve energy, unlike symplectic integrators, assuming convergence in the Hamiltonian {\citep[Section IX.8]{hair06}}.

\section{Acknowledgements}
We thank Walter Dehnen for stimulating discussions {and detailed comments on the manuscript, the anonymous referee for comments that strengthened the paper, and Ernst Hairer for a helpful discussion.}

\appendix
\section{Symplecticity of Implicit Midpoint rule}
\label{sec:symp}

The Jacobian is
\begin{equation}\label{Jacobian}
  \bm{S}\equiv\frac{\partial{\bm y'}}{\partial{\bm y}}\ .
\end{equation}
For the ordering $\v{y} = (\v{q},\v{p})$ (which we can choose without loss of generality), the symplectic condition is a matrix equation with $2n^2 + n$ independent constraints,
\begin{equation}\label{SympJacob}
  \bm{J}=\bm{SJS}^\dagger\ ,\ \ \bm{J}\equiv\begin{pmatrix} \bm{0} & \bm{I}_n\cr -\bm{I}_n & \bm{0}\end{pmatrix}\ .
\end{equation}
Here, $\v{y}$ has $2n$ elements while $\bm{I}_n$ is the $n\times n$ identity matrix.
Let $\bm{I} \equiv \bm{I}_{2n}$.  Differentiating the implicit midpoint rule, (\ref{eq:mid}), with respect to $\v{y}$ gives,
\begin{eqnarray}\label{Jacmid1}
  \bm{S}&=&\bm{I}+h\frac{\partial}{\partial{\bm y}}{\bm f}\left(\frac{{\bm y}
    +{\bm y'}}{2}\right)\nonumber\\
    &=&\bm{I}+\frac{1}{2}h\bm{JH}\left(\bm{I}+\bm{S}\right)\ ,\ \ 
    \bm{H}\equiv{\bm\partial}{\bm\partial}H_0
\end{eqnarray}
where $H_0$ is the original Hamiltonian and $\bm{H}$ is its Hessian matrix {evaluated at the midpoint}. We can solve for $S$ to get,
\begin{equation}\label{Jacmid}
  \bm{S}=(\bm{I}-\bm{A})^{-1}(\bm{I}+\bm{A})\ ,\ \ \bm{A}\equiv\frac{1}{2}h\bm{JH}\ .
\end{equation}
Next, note that,
\begin{equation}\label{SAident2}
  (\bm{I}-\bm{A})\bm{J}(\bm{I}-\bm{A}^\dagger)=(\bm{I}+\bm{A})\bm{J}(\bm{I}+\bm{A}^\dagger).
 \end{equation}
From this we find that,
\begin{equation}
\bm{SJS}^\dagger = (\bm{I}-\bm{A})^{-1}(\bm{I}+\bm{A}) \bm{J} (\bm{I} + \bm{A}^\dagger) (\bm{I} - \bm{A}^\dagger)^{-1} = \bm{J},
\end{equation}
which implies the midpoint rule is symplectic.

\section{Modified differential equation for trapezoidal and implicit midpoint rule}\label{sec:MDEMT}

We derive the MDEs for the symmetric trapezoidal rule in order to understand its error properties and whether it conserves energy.  Although Hamilton's equations split the configuration space into coordinates and momenta, the numerical integration algorithms need not do so. For RK-methods, in particular, the update rules depend on scalar operators formed from ${\bm f}$ and ${\bm\partial}\equiv\partial/\partial{\bm y}$. These operators are defined so that
\begin{equation}\label{fnexp}
  {\bm f}_n=\sum_{m=0}^{M_{n-1}}f_{nm}\hat D_{nm}{\bm f}\ ,
\end{equation}
where ${\bm f}_n({\bm y})$ is the $n$th-order contribution to the modified differential equation, $f_{nm}$ are constants, and $\hat D_{nm}$ are scalar differential operators. {$f_{n m}$ is not a component of $\v{f}_n$.  $M_n$ is the number of unlabeled rooted trees with $n$ nodes \citep[Table 2.1]{Haireretal93}}  Note that $\hat D_{nm}{\bm f}$ provide a basis for the Hilbert space of ${\bm F}$. Expansion (\ref{fnexp}) represents the function $F({\bm y},h)$ by a set of constants $f_{nm}$.

We now show how to obtain all such operators recursively in powers of $h$. At first order $(n=1)$ there is only one operator,
\begin{equation}\label{D1def}
  \hat D_{10}\equiv{\bm f}\cdot\frac{\partial}{\partial{\bm y}}\equiv
  f_i\partial_i.
\end{equation}
The implied summation of $i$ is from 1 to $2n$. {The subscripts $f_i$ indicate components of $\v{f}$ in \eqref{sysode} in equations \eqref{D1def}--\eqref{ftrap}, they are not the indices of \eqref{TaylorFG}.}    There is no other scalar operator that can be formed from ${\bm f}$ and $\partial/\partial{\bm y}$ that has units of ${\bm f}/h$, hence $M_1=0$. In equations (\ref{Dexp})--(\ref{gf}), $\hat D_{10}$ was written as $D_0$.

At second order ($n=2$), there are two linearly independent scalar operators with the correct units:
\begin{equation}\label{D2def}
  \hat D_{20}\equiv (\hat D_{10}f_i)\partial_i\ ,\ 
  \hat D_{21}\equiv f_if_j\partial_i\partial_j\ .
\end{equation}
Note that $\hat D_{10}^2=\hat D_{20}+\hat D_{21}$. We exclude $f_if_i\partial_j\partial_j$ and similar operators, even though they are scalars, because they do not arise in the series expansion of Runge{--}Kutta methods. At third order, there are four operators:
\begin{equation}\label{D3def}
\begin{aligned}
  \hat D_{30} &\equiv (\hat D_{20}f_i)\partial_i\ ,\ \ 
  \hat D_{31}\equiv (\hat D_{21}f_i)\partial_i\ ,\ \ 
  \hat D_{32}\equiv f_i(\hat D_{10}f_j)\partial_i\partial_j\ ,\ \  \\
  \hat D_{33} & \equiv f_if_jf_k\partial_i\partial_j\partial_k\ .
  \end{aligned}
\end{equation}

At fourth order, there are $9$ operators:
\begin{eqnarray}\label{D4def}
  &&\hat D_{40}\equiv (\hat D_{30}f_i)\partial_i\ ,\ \ 
  \hat D_{41}\equiv (\hat D_{31}f_i)\partial_i\ ,\ \ 
  \hat D_{42}\equiv (\hat D_{32}f_i)\partial_i\ ,\ \ 
  \hat D_{43}\equiv (\hat D_{33}f_i)\partial_i\ ,\nonumber\\
  &&\hat D_{44}\equiv f_i(\hat D_{20}f_j)\partial_i\partial_j\ ,\ \ 
  \hat D_{45}\equiv f_i(\hat D_{21}f_j)\partial_i\partial_j\ ,\ \ 
  \hat D_{46}\equiv (\hat D_{10}f_i)(\hat D_{10}f_j)\partial_i\partial_j\ ,
  \nonumber\\
  &&
  \hat D_{47}\equiv f_if_j(\hat D_{10}f_k)\partial_i\partial_j\partial_k\ ,\ \ 
  \hat D_{48}\equiv f_if_jf_kf_l\partial_i\partial_j\partial_k\partial_l\ .
\end{eqnarray}
The pattern becomes clear: at order $n$, the first $M_{n-1}$ operators are formed from the operators of order $(n-1)$ acting on $f_i$ combined with $\partial_i$ while the remaining operators are formed from operators of order $n-2,n-3,\ldots,0$ and additional derivative operators. At fifth order there are a total of $M_5=20$ operators; the first 9 are $\hat D_{5m}=(\hat D_{4m}f_i)\partial_i$. Note that the units of $D_{nm}$ are $h^{-n}$.

Using equations (\ref{TaylorFG}) and (\ref{fnexp}), the time evolution operator is now
\begin{equation}\label{Dhat}
  \hat D={\bm F}\cdot{\bm\partial}=\sum_{n=0}^\infty h^n
    \sum_{m=0}^{M_n}f_{nm}(\hat D_{nm}{\bm f})\cdot{\bm\partial}
  =\sum_{n=0}^\infty h^n\sum_{m=0}^{M_n}f_{nm}
    \hat D_{n+1,m}\ .
\end{equation}

For the trapezoidal rule, \eqref{eq:trap}, Taylor expanding ${\bm f}({\bm y}_1)$ about ${\bm y}_0$ gives,
\begin{equation}
\label{gtrap}
\begin{aligned}
  {\bm g}_0&={\bm f} \\
  {\bm g}_1&=\frac{1}{2}D_0{\bm f}=\frac{1}{2}\hat D_{10}{\bm f}
    \nonumber\\ 
  {\bm g}_2&=\frac{1}{4}D_0^2{\bm f}=\frac{1}{4}(\hat D_{20}+
    \hat D_{21}){\bm f}\nonumber\\ 
  {\bm g}_3&=\frac{1}{12}D_0^3{\bm f}+\frac{1}{24}(D_0^2f_j)
    (\partial_j{\bm f})=\frac{1}{8}\left(\hat D_{30}+\hat D_{31}+2\hat
    D_{32}+\frac{2}{3}\hat D_{33}\right){\bm f} \\
  {\bm g}_4&=\frac{1}{48}\left\{D_0^4{\bm f}+(D_0^2f_j)
    \left[\partial_j(D_0{\bm f})\right]+D_0\left[(D_0^2f_j)
    (\partial_j{\bm f})\right]\right\} \\
    &=\frac{1}{16}\left(\hat D_{40}+\hat D_{41}+2\hat D_{42}  +\frac{2}{3}\hat D_{43} +2\hat D_{44}+2\hat D_{45}+\hat D_{46} \right. \\
    & \left. 
      +2\hat D_{47}+\frac{1}{3}\hat D_{48}\right){\bm f}.\quad
\end{aligned}
\end{equation}
Substituting into (\ref{gf}) and solving for ${\bm f}_n$ gives
\begin{equation}\label{ftrap}
\begin{aligned}
 & {\bm f}_0={\bm f}\nonumber\\
  &{\bm f}_1=0\nonumber\\ 
& {\bm f}_2=\frac{1}{12}D_0^2{\bm f}=\frac{1}{12}(\hat D_{20}+
    \hat D_{21}){\bm f}\nonumber\\
  &{\bm f}_3=0\nonumber\\
  &{\bm f}_4=-\frac{1}{720}D_0^4{\bm f}+\frac{1}{144}\left\{
    (D_0^2f_j)\left[\partial_j(D_0{\bm f})\right]+D_0\left[(D_0^2f_j)
    (\partial_j{\bm f})\right]\right\}\nonumber\\
  &=\frac{1}{240} \times \\
  &\left(3\hat D_{40}+3\hat D_{41}+4\hat D_{42}+
    \frac{4}{3}\hat D_{43}+2\hat D_{44}+2\hat D_{45}-\hat D_{46}
    -2\hat D_{47}-\frac{1}{3} \hat D_{48}\right){\bm f}\nonumber\\
    \end{aligned}
\end{equation}

For the implicit midpoint rule, \eqref{eq:mid},
\begin{equation}
\label{gmid}
\begin{aligned}
  {\bm g}_0=&{\bm f}\nonumber\\
  {\bm g}_1=&\frac{1}{2}\hat D_{10}{\bm f}\nonumber\\ 
  {\bm g}_2=&\frac{1}{4}\left(\hat D_{20}+\frac{1}{2}\hat D_{21}
    \right){\bm f}\nonumber\\ 
  {\bm g}_3=&\frac{1}{8}\left(\hat D_{30}+\frac{1}{2}\hat D_{31}
    +\hat D_{32}+\frac{1}{6}\hat D_{33}\right){\bm f}\nonumber\\
  {\bm g}_4=&\frac{1}{16}\left(\hat D_{40}+\hat D_{42}+
    \hat D_{44}+\frac{1}{2}(\hat D_{41}+\hat D_{45}+\hat D_{46}
    +\hat D_{47})+\frac{1}{6}\hat D_{43}+\frac{1}{24}\hat D_{48}
    \right) \v{f}
\end{aligned}
\end{equation}
which leads to 
\begin{equation}
\label{fmid}
\begin{aligned}
 & {\bm f}_0={\bm f}\ ,\ \ {\bm f}_1=0\ ,\ \ 
  {\bm f}_2=\frac{1}{12}\left(\hat D_{20}-\frac{1}{2}\hat D_{21}
    \right){\bm f}\ ,\ \ {\bm f}_3=0\ ,\\
  &{\bm f}_4= \frac{1}{480} \times \\
  &\left(6(\hat D_{40}-\hat D_{44})
    +\hat D_{41}-2\hat D_{42}-\hat D_{45}+3\hat D_{46}+\hat D_{47}
    +\frac{7}{12}(-4\hat D_{43}+\hat D_{48})\right) \v{f}\ .
\end{aligned}
\end{equation}
To derive the Hamiltonian {for the} midpoint rule to fourth order, we use the procedure of Appendix \ref{sec2}. It is
\begin{equation}\label{Hammid}
  \tilde H=H-\frac{h^2}{24}(\hat D_{21}H)+\frac{h^4}{480}\left(
    3\hat D_{40}+\hat D_{41}-\frac{7}{12}\hat D_{43}\right)H+O(h^6)\ ,
\end{equation}
where $H$ is the original Hamiltonian.


\section{Runge{--}Kutta methods and energy conservation}\label{sec2}

We now obtain some general results concerning energy conservation for Runge{--}Kutta methods based on Hamiltonian systems. In this section we do not assume that energy conservation implies canonical transformation, even though the reverse is true (canonical transformation implies the existence of a local Hamiltonian, hence energy conservation for a time-independent Hamiltonian).

We consider conservative systems, for which equations (\ref{sysode}) take the form of Hamilton's equations,
\begin{equation}\label{Hamilton}
  \frac{dq^I}{dt}=\frac{\partial H}{\partial p_I}\equiv H^I\ ,\ \ 
  \frac{dp_I}{dt}=-\frac{\partial H}{\partial q^I}\equiv-H_I\ ,\ \ 
  H=H({\bm q},{\bm p})\ .
\end{equation}
{We use superscript and lowerscript indices to distinguish derivatives with respect to coordinates and momenta.  Einstein summation convention is also used.  We use these notations only in this Appendix to simplify results.} For a configuration space of {$n$} coordinates and {$n$} momenta, indices range from $1$ to {$n$}.

We now ask under what conditions RK methods applied to a conservative Hamiltonian system have a conserved energy
\begin{equation}\label{Eexp}
  E({\bm y},h)=H_0({\bm y})+hH_1({\bm y})+h^2H_2({\bm y})
    +h^3H_3({\bm y})+h^4H_4({\bm y})+\cdots
\end{equation}
such that $E$ is constant for solutions of the modified differential equation. (Note that we are not requiring the integrator to be symplectic; the relationship between symplectic and energy-conserving integrators will be clarified later.) In other words, the solutions must obey $\hat DE=F_i\partial_iE=0$. Applying (\ref{Dhat}) to (\ref{Eexp})  gives
\begin{equation}\label{Econsn}
  \sum_{k=0}^n\sum_{m=0}^{M_{k-1}}f_{km}\hat D_{k+1,m}H_{n-k}=0
\end{equation}
for all $n\ge0$, with $M_0=f_{00}=1$.

Let's examine this order by order. For $n=0$, equation (\ref{Econsn}) is automatically satisfied because $\hat D_{10}H_0=\hat D_{10}H=\{H,H\}=0$: we are numerically integrating a Hamiltonian system. For $n=1$, energy conservation requires that there exist a $H_1({\bm y})$ satisfying
\begin{equation}\label{Econs1}
  \hat D_{10}H_1=-f_{10}\hat D_{20}H_0
\end{equation}
For a{n} RK method, $H_n$ can be formed only from $H_0$ and scalar derivative operators $D_{nm}$. For $n=1$, there is only one such operator, $\hat D_{10}$, and $\hat D_{10}H_0=0$. {Therefore, {\it no first-order RK method has a conserved energy.} Examples are the explicit and implicit Euler methods, which usually exhibit a growth in the absolute value of the energy error that is linear in time.  This behavior is explained with other numerical analysis.  The only possibility that allows a conserved energy is $f_{10}=0$, i.e. ${\bm f}_1=0$ and the integration method is at least second order.

As with the function ${\bm F}({\bm y},h)$, for a{n} RK-method we must represent $E({\bm y},h)$ using scalar operators and the unique scalar function corresponding to ${\bm f}$, namely $H_0$. Thus, in equation (\ref{Econsn}) we write
\begin{equation}\label{Hexp}
  H_n=\sum_{m=0}^{M_{n-1}}e_{nm}\hat D_{nm}H_0\ .
\end{equation}
The following results are obtained (after much algebra) from equations (\ref{D1def})--(\ref{D4def}): 
\begin{equation}\label{Hres1}
\begin{aligned}
  \hat D_{10}H_0&=0\ ,\ \hat D_{20}H_0=-\hat D_{21}H_0\ ,\ 
  \hat D_{30}H_0=\hat D_{32}H_0=0\ ,\  \\
  \hat D_{31}H_0&=-\hat D_{33}H_0\ ,\\
  \hat D_{40}H_0&=-\hat D_{44}H_0=D_{46}H_0=(\hat D_{10}H^I)
    (\hat D_{20}H_I)-(\hat D_{10}H_I)(\hat D_{20}H^I)\ ,\\ 
  \hat D_{41}H_0&=-\hat D_{42}H_0=-\hat D_{45}H_0
    =\hat D_{47}H_0 \\
    &=(\hat D_{10}H^I)(\hat D_{21}H_I)
      -(\hat D_{10}H_I)(\hat D_{21}H^I)\ ,\\
  \hat D_{43}H_0&=-\hat D_{48}H_0\ ,\\
  \hat D_{50}H_0&=\hat D_{55}H_0=0\ ,\ 
  \hat D_{51}H_0=\hat D_{54}H_0\\
  &=(\hat D_{20}H_I)(\hat D_{21}H^I)
    -(\hat D_{20}H^I)(\hat D_{21}H_I)\ ,\\
  \hat D_{52}H_0&=-D_{56}H_0=(\hat D_{10}H^I)(\hat D_{32}H_I)-
    (\hat D_{10}H_I)(\hat D_{32}H^I)\ ,\\
  \hat D_{53}H_0&=-\hat D_{57}H_0=(\hat D_{10}H^I)(\hat D_{33}H_I)
    -(\hat D_{10}H_I)(\hat D_{33}H^I)\ .
    \end{aligned}
\end{equation}
On account of these results, many of the dimensionless coefficients $e_{nm}$ can be set to zero without loss of generality, so that
\begin{equation}\label{Eexp1}
\begin{aligned}
  E({\bm y},h)&=H_0+h^2e_{21}\hat D_{21}H_0+h^3e_{31}\hat D_{31}
   H_0 \\
   &+h^4\left(e_{40}\hat D_{40}+e_{41}\hat D_{41}+e_{43}
   \hat D_{43}\right)H_0 +O(h^5)\ .
   \end{aligned}
\end{equation}
The task is now to find expressions for the $e_{nm}$ in terms of the $f_{nm}$, as well as any conditions on the $f_{nm}$ that must be satisfied in order to have energy conservation.

The following identities are also useful:
\begin{equation}\label{Hres2}
\begin{aligned}
  \hat D_{10}\hat D_{21}H_0 &=-\hat D_{31}H_0\ ,\ 
  \hat D_{10}\hat D_{31}H_0=(-3\hat D_{41}+\hat D_{43})H_0\ ,
    \nonumber\\
  \hat D_{30}\hat D_{20}H_0&=-\hat D_{30}\hat D_{21}H_0
  =\frac{1}{2}\hat D_{31}\hat D_{20}H_0=-\frac{1}{2}\hat D_{31}
  \hat D_{21}H_0=\hat D_{51}H_0\ ,\nonumber\\
  \hat D_{10}\hat D_{40}H_0&=(2\hat D_{51}+\hat D_{52})H_0\ ,\ \ 
  \hat D_{10}\hat D_{41}H_0=(-\hat D_{51}+2\hat D_{52}
    +\hat D_{53})H_0\ ,\nonumber\\
  \hat D_{10}\hat D_{43}H_0&=(\hat D_{58}-4\hat D_{53})H_0
  \end{aligned}
\end{equation}
Combining these results gives the conditions for energy conservation up to fourth order:
\begin{eqnarray}\label{Econs1234}
  O(h^1):&&f_{10}=0\nonumber\\  
  O(h^2):&&e_{21}=f_{21}\nonumber\\
  O(h^3):&&f_{30}=0\ ,\ f_{31}-f_{32}+3f_{33}=0\ ,\ e_{31}=-f_{33}
    \nonumber\\
  O(h^4):&&f_{41}+f_{44}-2(f_{42}-f_{46})+5(f_{43}-f_{47}+4f_{48})
    -f_{21}(f_{20}+2f_{21})=0\ ,\nonumber\\
  &&e_{40}=-f_{42}+f_{46}+2(f_{43}-f_{47}+4f_{48})\ ,\nonumber\\
  &&e_{41}=-f_{43}+f_{47}-4f_{48}\ ,\ e_{43}=-f_{48}\ .
\end{eqnarray}
The equations involving no $e_{nm}$ are constraints on the numerical method in order that it have a conserved energy. At second order, there is no constraint: every second-order RK method has a conserved energy to second order, regardless whether the method is symplectic.  For example, the explicit midpoint method typically shows linear growth in the absolute value of the energy.  It is a second order RK method, but the slope of the linear drift scales as $h^3$, as we can check. At third order, there are two constraints on the four coefficients $f_{3m}$, so that most third-order $y$-methods do not have third-order energy conservation property. At fourth order, there is one constraint on the nine coefficients $f_{4m}$ in order that a conserved energy result.

Kutta's third order method violates energy conservation at third order, while the classic Runge{--}Kutta fourth order method violates energy conservation at fourth order.  

We will see in Appendix \ref{sec3} that symplectic methods have additional constraints beyond those given above. Symmetric integrators are purely even in $h$, so that $f_{nm}=0$ for odd $n$. Not all symmetric integrators have a conserved energy, but all symplectic ones do. Thus, the set of symplectic integrators is a subset of the set of energy-conserving ones, and the set of symmetric integrators overlaps with both.  {Recall for non-adaptive one-step methods, time-symmetry and time-reversibility are equivalent.}  

\section{Symplectic Runge{--}Kutta-methods}\label{sec3}

Symplectic integrators are ones for which the mapping ${\bm y}_0\to{\bm y}_1$ is a canonical transformation. In this case the modified differential equation (\ref{MDE}) is equivalent to Hamilton's equations (\ref{Hamilton}) with modified Hamiltonian $H({\bm y},h)$. The modified Hamiltonian is expanded in power series exactly the same as $E({\bm y},h)$ in equation (\ref{Eexp}); we will use the same coefficients, with the expectation that requiring the integrator to be symplectic will yield different constraints than equations (\ref{Econs1234}). Enforcing symplecticity requires using the following identities,
\begin{equation}\label{Hres3}
\begin{aligned}
  \partial_i(\hat D_{20}H)&=(2\hat D_{20}-\hat D_{21})H_i\ ,\ 
  \partial_i(\hat D_{31}H)=(3\hat D_{31}-\hat D_{33})H_i\ ,\nonumber\\
  \partial_i(\hat D_{40}H)&=\left[2(\hat D_{40}-\hat D_{44})
    +\hat D_{46}\right]H_i\ ,\ 
  \partial_i(\hat D_{43}H)=(4\hat D_{43}-\hat D_{48})H_i\ ,\nonumber\\
  \partial_i(\hat D_{41}H)&=(\hat D_{41}-2\hat D_{42}-\hat D_{45}
    +\hat D_{47})H_i\ .
    \end{aligned}
\end{equation}
Applying these gives the following conditions for symplecticity of RK-integrators, up to fourth order in $h$:
\begin{eqnarray}\label{symp1234}
  O(h^1):&&f_{10}=0\nonumber\\  
  O(h^2):&&f_{20}=-2f_{21},\ e_{21}=f_{21}\nonumber\\
  O(h^3):&&f_{30}=f_{32}=0,\ f_{31}=-3f_{33},\ 
    e_{31}=\frac{1}{3}f_{31}\nonumber\\
  O(h^4):&&f_{40}=-f_{44}=2f_{46},\ f_{41}=-\frac{1}{2}f_{42}
    =-f_{45}=f_{47},\ f_{43}=-4f_{48},\nonumber\\
  &&e_{40}=\frac{1}{2}f_{40},\ e_{41}=f_{41},\ 
    e_{43}=\frac{1}{4}f_{43}\ .
\end{eqnarray}
Notice that these conditions include, but are stronger than, the energy-conserving conditions (\ref{Econs1234}). Symplectic integrators for an autonomous Hamiltonian system are always energy-conserving. However, the set of energy-conserving integrators is larger: up to fourth order, there are energy-conserving Runge{--}Kutta methods that are not symplectic, such as Lobatto IIIA {(but note Lobatto IIIA does not conserve energy at higher orders according to \cite{FHP04}).  Lobatto IIIB, to fourth order, is neither symplectic nor conserves energy}. There also exist third-order symplectic integrators, which are not symmetric.

\section{General Runge{--}Kutta integrators}\label{sec5}

The general $s$-stage Runge{--}Kutta method can be written
\begin{equation}\label{RKgen}
  {\bm y}'={\bm y}+h\sum_{i=1}^sb_i{\bm k}_i\ ,\ \ 
  {\bm k}_i={\bm f}({\bm y}+h{\bm q}_i)\ ,\ \ 
  {\bm q}_i\equiv[a_{ij}{\bm k}_j]\equiv\sum_{j=1}^sa_{ij}{\bm k}_j\ .
\end{equation}
Square brackets indicate a sum over the repeated indices inside the sum, e.g.
\begin{equation}\label{bracketnot}
  [a_{ij}c_j^2]\equiv\sum_{j=1}^sa_{ij}c_j^2\ ,\ \ 
  [a_{ij}a_{jk}c_k]\equiv\sum_{j=1}^s\sum_{k=1}^sa_{ij}a_{jk}c_k\ .
\end{equation}
We also define
\begin{equation}\label{cdef}
  c_i\equiv\sum_{j=1}^sa_{ij}\ .
\end{equation}
The equation (\ref{RKgen}) for ${\bm k}_i$ is recursive. Expanding in power series in $h$ gives ${\bm k}_i=\hat K_i{\bm f}$, where the propagator is
\begin{equation}\label{RKopser}
\begin{aligned}
  \hat K_i&=1+hc_i\hat D_{10}+h^2[a_{ij}c_j]\hat D_{20}
     +\frac{1}{2}h^2c_i^2\hat D_{21}+h^3[a_{ij}a_{jk}c_k]\hat
      D_{30}\\
      &+\frac{1}{2}h^3[a_{ij}c_j^2]\hat D_{31}
  +h^3c_i[a_{ij}c_j]\hat D_{32}+\frac{1}{6}h^3c_i^3
    \hat D_{33}+h^4[a_{ij}a_{jk}a_{kl}c_l]\hat D_{40} \\
    &+\frac{1}{2}
    h^4[a_{ij}a_{jk}c_k^2]\hat D_{41}
  +h^4[a_{ij}a_{jk}c_jc_k]\hat D_{42}+\frac{1}{6}h^4[a_{ij}c_j^3]
    \hat D_{43} \\
    &+h^4c_i[a_{ij}a_{jk}c_k]\hat D_{44}+\frac{1}{2}h^4
    c_i[a_{ij}c_j^2]\hat D_{45}+\frac{1}{2}h^4[a_{ij}c_j]^2\hat D_{46}\\
    &+\frac{1}{2}h^4c_i^2
    [a_{ij}c_j]\hat D_{47}+\frac{1}{24}h^4c_i^4\hat D_{48}+O(h^5)
    \,\qquad\qquad\qquad\qquad\quad\ .
    \end{aligned}
\end{equation}
The integrator method is now
\begin{equation}\label{GRK}
  {\bm G}=\sum_{i=1}^sb_i\hat K_i{\bm f}
\end{equation}
This can be used for various methods to check the order of an integrator, its energy conservation properties, and its symplecticity, order by order.

Symmetric integrators are a special class of integrators for which equation (\ref{reverse}) holds.  For $s=1$, the implicit midpoint method is the only symmetric Runge{--}Kutta integrator. For $s=2$, the general class is defined by two parameters $(a_{11},a_{12})$ through the Runge{--}Kutta matrix
\begin{equation}\label{symm2}
  A_{\rm symm2}=\begin{pmatrix}a_{11} & a_{12}\\
  							\frac{1}{2}-a_{12}&\frac{1}{2}-a_{11}\\
  						  \end{pmatrix}\ ,\ \ 
  b_{\rm symm2}=\begin{pmatrix}\frac{1}{2} & \frac{1}{2} \\
  						  \end{pmatrix}\ .
\end{equation}
{Elements of $A_{\rm symm2}$ and $b_{\rm symm2}$ are, respecitvely, the $a_{i j}$ and $b_i$ from eq. \eqref{eq:RK}.}  These integrators are all at least second order because symmetry implies ${\bm f}_1={\bm f}_3=0$. They are not, in general, symplectic. There is one choice of $(a_{11},a_{12})$ for which the integrator is fourth order and symplectic (symplecticity to all orders is proved elsewhere), namely the Gauss{--}Legendre case
\begin{equation}\label{GL2}
  a_{11}=\frac{1}{4}\ ,\ \ a_{12}=\frac{1}{4}-\frac{\sqrt{3}}{6}\ .
\end{equation}
The general $s=3$ symmetric integrator has Runge{--}Kutta matrix and weight vector
\begin{equation}\label{symm3}
  A_{\rm symm3}=\begin{pmatrix}a_{11} & a_{12} & a_{13}\\
  							a_{21} & \frac{1}{2}b_2& b_1-a_{21}\\
  							b_1-a_{13} & b_2-a_{12} & b_1-a_{11}\\
  						  \end{pmatrix}\ ,\ \ 
  b_{\rm symm3}=\begin{pmatrix}b_1 & b_2 & b_1\\
  						  \end{pmatrix}
\end{equation}
with $b_2=1-2b_1$. In all cases this integrator is at least second order; in general it is not symplectic. The integrator is at least fourth order if the parameters obey the following two relations:
\begin{equation}\label{symm4}
\begin{aligned}
  a_{11}+a_{12}+a_{13}&=\frac{1}{2}\pm(24b_1)^{-1/2}\ ,\ \  \\
  b_1(a_{11}-a_{13})+b_2\left(a_{21}-\frac{1}{2}b_2\right)&=\mp\left(
    \frac{b_1}{24}\right)^{1/2}\ .
    \end{aligned}
\end{equation} 
There is one choice of parameters for which the integrator is sixth order and (at least to sixth order) symplectic, namely the Gauss{--}Legendre case
\begin{equation}\label{GL3}
\begin{aligned}
  a_{11}&=\frac{5}{36}\ ,\ \ a_{12}=\frac{2}{9}-\frac{\sqrt{15}}{15}\ ,\ \ 
    a_{33}=\frac{5}{36}-\frac{\sqrt{15}}{30}\ ,\ \ 
    a_{21}=\frac{5}{36}+\frac{\sqrt{15}}{24}\ ,\ \  \\
    b_1&=\frac{5}{18}\ .
    \end{aligned}
\end{equation}
Note that Gauss{--}Legendre methods have twice the order of truncation error expected from a naive count of function evaluations (e.g., 6 versus 3). In the context of Runge{--}Kutta methods this arises naturally because of symmetry: all odd terms vanish in the truncation error of the modified differential equation.
\begin{table}
\caption{Properties of implicit Runge{--}Kutta integrators.  For various methods, we state the number of stages ($s$ in eq. \eqref{eq:RK}), the order, and whether to all orders the methods are symmetric, energy conserving, and symplectic.}
\centering
\begin{tabular}{lccccc}
\hline\hline
Method & Stages & Order & Symm. & Econs & Symp\\
\hline
Midpoint & 1 & 2 & yes & yes & yes \\
\hline
Trapezoidal & 2 & 2 & yes & yes & no\\
Symmetric & 2 & $\ge2$ & yes & no & no\\
Gauss{--}Legendre & 2 & 4 & yes & yes & yes\\
\hline
Lobatto IIIA & 3 & 4 & yes & no & no\\
Lobatto IIIB & 3 & 4 & yes & no & no\\
Symmetric & 3 & $\ge$ 2 & yes & no & no\\
Gauss{--}Legendre & 3 & 6 & yes & yes & yes\\
\hline
\label{table:RK}
\end{tabular}
\end{table}
Table \ref{table:RK} summarizes various {properties of } implicit Runge{--}Kutta integrators.

It has been shown that the general conditions for symplecticity of any Runge{--}Kutta integrator are \citep[Chapter VI]{hair06}
\begin{equation}\label{SympRK}
  b_ia_{ij}+b_ja_{ij}=b_ib_j\ \hbox{for all $i,j$ such that $1\le i,j\le s$.}
\end{equation}
These conditions are satisfied by Gauss{--}Legendre integrator{s} but not by Lobatto III {integrators}. \cite[Chapter VI]{hair06} shows that Gauss-collocation methods (including the Gauss{--}Legendre methods above) are symplectic. 
\bibliographystyle{mnras}
\bibliography{refs}
\end{document}